\documentclass[journal,10pt]{IEEEtran}

\IEEEoverridecommandlockouts
\usepackage{cite}
\usepackage{caption}
\usepackage{amsmath,amssymb,amsfonts}
\usepackage{algorithmic}
\usepackage{graphicx}
\usepackage{tabularx}
\usepackage{inputenc}
\usepackage{textcomp}
\usepackage{xcolor}
\usepackage{stfloats}
\usepackage{multirow}
\def\BibTeX{{\rm B\kern-.05em{\sc i\kern-.025em b}\kern-.08em
    T\kern-.1667em\lower.7ex\hbox{E}\kern-.125emX}}
\begin{document}

\title{A General 3D Space-Time-Frequency Non-Stationary THz Channel Model for 6G Ultra-Massive MIMO Wireless Communication Systems}

\author{Jun Wang, \textit{Student Member, IEEE}, Cheng-Xiang Wang, \textit{Fellow, IEEE}, Jie Huang, \textit{Member, IEEE}, Haiming Wang, \textit{Member, IEEE}, and Xiqi Gao, \textit{Fellow, IEEE}
	\thanks{Manuscript received July 7, 2020; revised November 15, 2020 and February 20, 2021; accepted March 1, 2021. This work was supported by the National Key R\&D Program of China under Grant 2018YFB1801101, the National Natural Science Foundation of China (NSFC) under Grant 61960206006 and 61901109,  the Frontiers Science Center for Mobile Information Communication and Security, the High Level Innovation and Entrepreneurial Research Team Program in Jiangsu, the High Level Innovation and Entrepreneurial Talent Introduction Program in Jiangsu, the Research Fund of National Mobile Communications Research Laboratory, Southeast University, under Grant 2020B01 and Grant 2021B02, the Fundamental Research Funds for the Central Universities under Grant 2242020R30001, and the EU H2020 RISE TESTBED2 project under Grant~872172. \emph{(Corresponding author: Cheng-Xiang Wang.)}}
	\thanks{J. Wang, C.-X. Wang, J. Huang, and  X. Q. Gao are with the National Mobile Communications Research Laboratory, School of Information Science and Engineering, Southeast University, Nanjing, 210096, China, and also with the Purple Mountain Laboratories, Nanjing, 211111, China (email: {\{jun.wang, chxwang, j\_huang, xqgao\}@seu.edu.cn}). }
	\thanks{H. Wang is with the State Key Laboratory of Millimeter Wave, School of Information Science and Engineering, Southeast University, Nanjing, 210096, China, and also with the Purple Mountain Laboratories, Nanjing, 211111, China (email: hmwang@seu.edu.cn).}

}

\markboth{IEEE JOURNAL ON SELECTED AREAS IN COMMUNICATIONS,~Vol.~XX, No.~XX, MONTH~2021}%
{J. Wang \MakeLowercase{\textit{et al.}}: Bare Demo of IEEEtran.cls for IEEE Journals}

\maketitle

\begin{abstract}
In this paper, a novel three-dimensional (3D) space-time-frequency (STF) non-stationary geometry-based stochastic model (GBSM) is proposed for the sixth generation (6G) terahertz (THz) wireless communication systems. The proposed THz channel model is very general having the capability to capture different channel characteristics in multiple THz application scenarios such as indoor scenarios, device-to-device (D2D) communications, ultra-massive multiple-input multiple-output (MIMO) communications, and long traveling paths of users. Also, the generality of the proposed channel model is demonstrated by the fact that it can easily be reduced to different simplified channel models to fit specific scenarios by properly adjusting model parameters. The proposed general channel model takes into consideration the non-stationarities in space, time, and frequency domains caused by ultra-massive MIMO, long traveling paths, and large bandwidths of THz communications, respectively. Statistical properties of the proposed general THz channel model are investigated. The accuracy and generality of the proposed channel model are verified by comparing the simulation results of the relative angle spread and root mean square (RMS) delay spread with corresponding channel measurements.

\end{abstract}

\begin{IEEEkeywords}
6G wireless communication systems, THz channel model, ultra-massive MIMO, long traveling path, space-time-frequency non-stationarity. 

\end{IEEEkeywords}

\section{Introduction}

The terahertz (THz) band (0.3--10 THz) is currently being explored  for the sixth generation (6G) wireless communication systems\cite{5Gsurvey,WangCX1,J20_SCIS_XHYou_6G,WangCX2}.  THz band  has huge bandwidth to provide ultra-high transmission data rate in numerous wireless applications. Some specific applications such as holographic video conferencing in the indoor room are supported by terabit wireless local area networks (T-WLAN). Data center\cite{datacenter1} is a brand new scenario where thousands of devices are connected for cloud computing and storage. The interconnection of nanoscale machines in THz band strongly supports the Internet of nano things \cite{nanothings1}. In addition, the THz band can provide ultra-high speed links for intercore communication in wireless on-chip networks \cite{chip1} with  nano on-chip antennas.  In the THz communication,  high transmission and reflection loss limit the transmission distance and massive multiple-input multiple-output (MIMO) technology is often utilized  to compensate the  transmission loss. In the aim of  designing and evaluating THz communication systems more efficiently, an accurate and general channel model that can accurately  capture THz propagation characteristics of different scenarios is essential. However, the existing THz channel models only focus  on  specific scenarios.

In THz communication systems, a significant challenge is the fact that the phenomena of scattering and diffraction are quite different from lower frequency bands.   
The THz propagation mechanisms were studied in  \cite{RN308,RN356,RN197,RN364,RN346,RN394,RN369}. THz measurements of multiple reflection effects on different materials were introduced in \cite{RN308,RN356}.   Due to high reflection loss, high-order paths were very hard to detect resulting in the limited number of multipaths in THz band  according to the measurement in \cite{RN197}. The diffusely scattered propagation played an important role and was investigated in\cite{RN364,RN346,RN394,RN369}. In \cite{RN364},  the detected signal powers in all directions for different materials were measured and simulated. Frequency-dependent scattering was also measured and simulated in \cite{RN346,RN394}. The wavelength of THz waves is in the same order  with the roughness of some common materials. From these measurements and simulations, we found that The proportion  of diffusely scattered rays gradually increased when it comes to higher frequency. In addition, each specular reflected path is surrounded with multiple   diffusely scattered rays\cite{RN369}.

A number of channel models and measurements were investigated  for indoor THz communications\cite{RN207,RN229,RN194,RN195,RN139,RN506,C16_ICC_Khalid}. In~\cite{RN207}, a multi-path ray tracing channel model  for THz indoor communication  
was presented and validated with experiments. In \cite{RN229}, a three-dimension (3D) time-variant THz ray tracing channel model was investigated for dynamic environments.  In \cite{RN194}, a geometry-based stochastic THz indoor channel model considering the frequency dispersion was presented and verified by ray tracing. In \cite{RN195}, the authors investigated root mean square (RMS) delay spread and angular spread that were modeled by second order polynomial parameters for THz indoor communications. An indoor channel model based on  ray tracing considering atmospheric attenuation was proposed in \cite{RN139}. Antenna arrays were discussed for indoor THz communication systems in \cite{RN506}. A wideband   channel measurement for  indoor THz  wireless links was conducted between  260 GHz and 400 GHz\cite{C16_ICC_Khalid}.
Wireless data center networks in THz band were investigated in the literature\cite{datacenter1,datacenter2,datacenter3,datacenter4,datacenter5}. In addition, high speed transmission for THz wireless data center was promised with its  large bandwidth and lower interference \cite{datacenter6,datacenter7,datacenter8,datacenter9,datacenter10,datacenter11}.
A stochastic channel model was proposed for THz data center and simulated in \cite{datacenter12,datacenter13}.

In THz communication systems, ultra-massive MIMO technologies employing thousands of antennas are considered as one of the solutions to compensate the path loss and are expected to be utilized in 6G\cite{WangCX1,WangCX2}. 
Massive MIMO channel models and measurements were studied in \cite{RN321,RN322,massiveMIMO1,massiveMIMO2,massiveMIMO3,massiveMIMO4,massiveMIMO_bianji}.  
In~\cite{RN321}, a detailed survey of  massive MIMO channel models and measurements  were summarized. In \cite{RN322,massiveMIMO1,massiveMIMO2,massiveMIMO3,massiveMIMO_bianji}, the effect of spherical wavefront and clusters evolution along the time and array axis are considered, showing the spatial non-stationarity in massive MIMO systems.  Non-stationary massive MIMO channels by transformation of delay and angle of arrivals were studied in \cite{massiveMIMO4}. In \cite{RN161,HuangjieJSAC,HuangjieCSCI,HeRuisiTVT2018,J_JSAC14_Akdeniz}, millimeter wave (mmWave) band channel measurements for massive MIMO channels in different scenarios  were presented. A novel 3D geometry-based stochastic model (GBSM) based on homogeneous Poisson point process was proposed for mmWave channels\cite{HuangjieCSCI}. A GBSM for mobile-to-mobile scenarios for mmWave bands was presented in \cite{HeRuisiTVT2018}.

In general, the existing massive MIMO channel models have only considered characteristics of sub-6~GHz and mmWave bands, and are not suitable for THz communication systems because propagation mechanisms in THz bands are quite different from  lower frequencies. The  ray tracing based deterministic channel models are so complex and not suitable for THz communication systems design. Different from deterministic channel models, the GBSMs are more flexible and widely used in the 5G standardized channel models\cite{3GPP38901}. However, the existing THz GBSMs do not support mobility, ultra-massive MIMO, and frequency non-stationarity simultaneously and only focus on one specific scenario. In order to design and evaluate 6G wireless communication systems more efficiently,  in this paper, a general 3D THz GBSM considering space-time-frequency (STF) non-stationarity is proposed. The major contributions  of this paper are listed as follows. 

\begin{enumerate}
	\item A general STF non-stationary THz channel model for 6G ultra-massive MIMO wireless communication systems is presented. The proposed model can support different scenarios whose transmission distances range from tens of meters to a few centimeters. The proposed THz channel is very general to for different  specific  THz scenarios and applications with certain model parameters.
	
	\item The STF non-stationarities in space, time, and frequency domains  caused by ultra-massive MIMO, long traveling paths, and large bandwidths are considered. The evolution in STF domain based on birth-death process is presented. 
	
	\item The statistical properties  such as STF correlation function (STFCF), the stationary interval, and power spectrum density (PSD) are derived. The simulation results show good agreements with the corresponding measurements, illustrating the validity and generality of the proposed THz simulation model.

\end{enumerate}

The remainder of this paper is organized as follows. 
In Section~\uppercase\expandafter{\romannumeral2}, the  THz GBSM is described in detail. The channel impulse response (CIR) and the  parameters are  presented.   The STF evolution based on birth-death process is introduced. In Section~\uppercase\expandafter{\romannumeral3}, typical statistical properties of the proposed general THz channel model are derived.  In  Section~\uppercase\expandafter{\romannumeral4}, different statistical properties of the channel model  are simulated and compared with measurements. Finally, conclusions are drawn in Section~\uppercase\expandafter{\romannumeral5}.
\section{A General 3D Non-Stationary THz Channel Model}

\begin{figure*}[tb]
	\centerline{\includegraphics[width=0.9\textwidth]{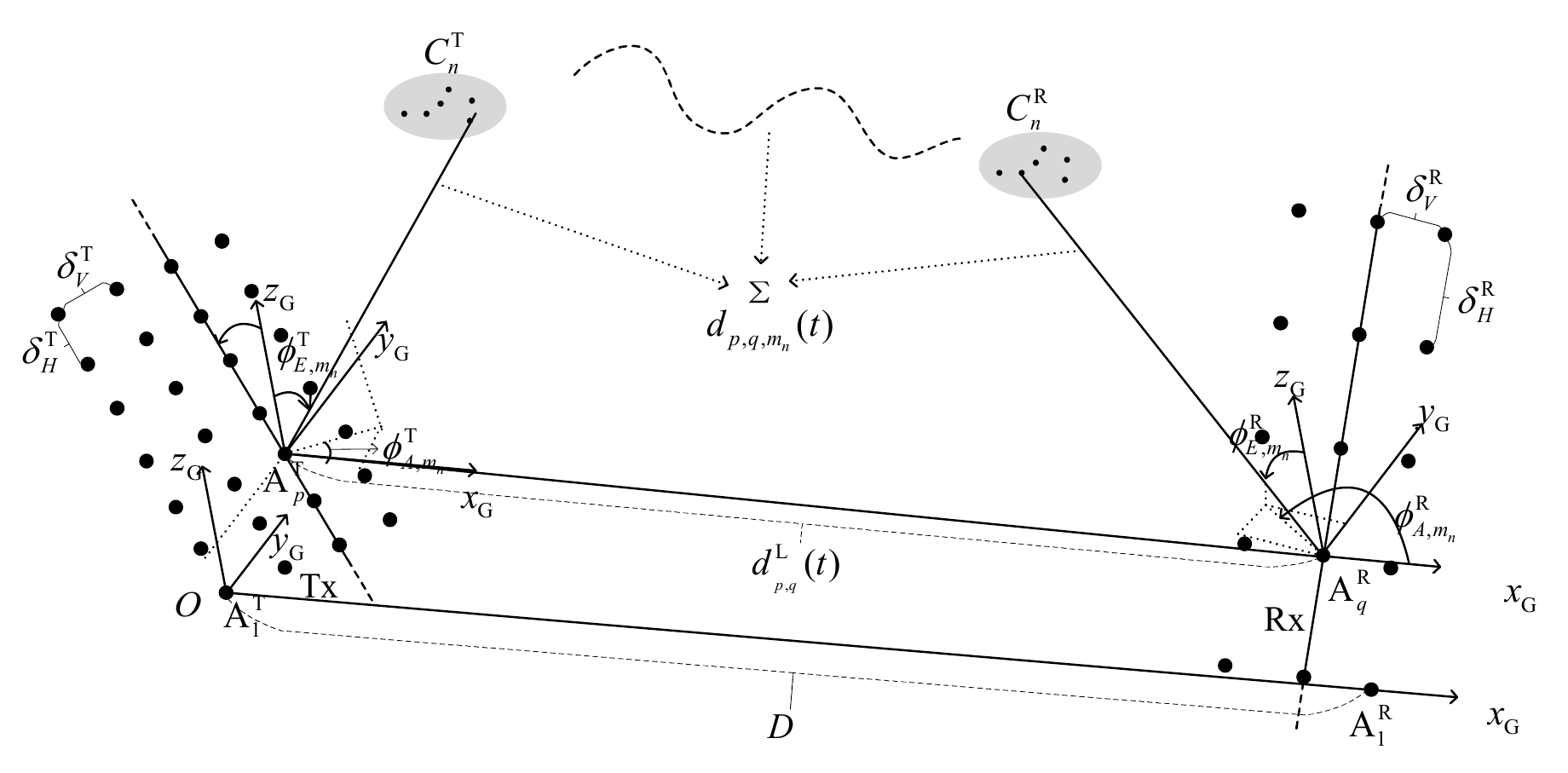}} 
	\caption{A 3D THz GBSM for massive MIMO communication systems.}
	\label{fig_1}
\end{figure*}

The diagram of THz  wireless communication system is shown in Fig. 1 where   uniform planar arrays (UPAs) are employed at the transmitter (Tx) and receiver (Rx) containing $M^{\text{T}}$ and $M^{\text{R}}$ elements at Tx and Rx, respectively. There are $M_V^{\text{T}}$ columns and $M_H^{\text{T}}$ rows in the Tx array. Similarly, $M_V^{\text{R}}$ columns and $M_H^{\text{R}}$ rows in the Rx array so that we have $M^{\text{T}} = M_V^{\text{T}}\times M_H^{\text{T}}$ and $M^{\text{R}} = M_V^{\text{R}}\times M_H^{\text{R}}$. {Note that non-uniform antenna arrays are also supported in the model with the knowledge of the geometric relationship of antenna elements.} As shown in Fig. 1, considering multi-bounce propagation, the $n$th path is represented by one-to-one pair clusters, i.e., $C^{\text{T}}_n$ and $C^{\text{R}}_n$ at the Tx and Rx side, respectively. In each path, 
The initial distance between the first element the Tx and  Rx is denoted as $D$. {The first element is the used as the reference to calculate other elements. It is a benchmark element in the array.} It should be noticed that $(x_\text{G}, y_\text{G}, z_\text{G})$ axes are established as the global coordinate system (GCS) whose  $x$ axis is oriented to the first element of the receive array. It should  be distinguished from the local coordinate systems (LCS)  when calculating 3D antenna pattern in the LCS.  The inter-element spacing in a row and in a column of Tx (Rx) are denoted as $\delta^\text{T(R)}_H$ and $\delta^\text{T(R)}_V$, respectively. The elevation and azimuth angels of the row (column) for Tx are denoted as $\beta^{\text{T}}_{V(H),E}$ and $\beta^{\text{T}}_{V(H),A}$. Let $\overrightarrow{A}_p^{\text{T}}$ and $\overrightarrow{A}_q^{\text{R}}$ denote the position vectors of $A_p^{\text{T}}$ and $A_q^{\text{R}}$ from the first element of Tx and Rx array, respectively. The $\overrightarrow{A}_p^{\text{T}}$ can be expressed as 
\begin{equation}
\begin{split}
&\overrightarrow{A}_p^{\text{T}}=\\
&p_H\delta^\text{T}_H
\left[
\begin{matrix}
\cos\beta^{\text{T}}_{H,E}\cos\beta^{\text{T}}_{H,A}
&
\cos\beta^{\text{T}}_{H,E}\sin\beta^{\text{T}}_{H,A}
&
\sin\beta^{\text{T}}_{H,E}
\end{matrix}
\right] 
\\
&+
p_V\delta^\text{T}_V
\left[
\begin{matrix}
\cos\beta^{\text{T}}_{V,E}\cos\beta^{\text{T}}_{V,A}
&
\cos\beta^{\text{T}}_{V,E}\sin\beta^{\text{T}}_{V,A}
&
\sin\beta^{\text{T}}_{V,E}
\end{matrix}
\right] 
\end{split}
\end{equation}
where $p_V$ and $p_H$ mean that the $p$th element is located in the $p_H$th row and $p_V$th column, we have that $p = (p_H-1)\times M_V^{\text{T}}+p_V$. For Rx array the vector $\overrightarrow{A}_q^{\text{R}}$ can be calculated in a similar way.

\begin{table*}[t]
	\centering
	\setlength{\belowcaptionskip}{0.2cm}
	\caption{Definitions of main parameters for the proposed THz  channel model.}
	\resizebox{\textwidth}{!}{
	\begin{tabular}{|l|l|}			
		\hline
		\textbf{Parameters}&\textbf{Definitions}\\
		\hline
		$\delta^\text{T(R)}_V$,$\delta^\text{T(R)}_H$&Vertical and horizontal inter-element spacing plane of Tx (Rx), respectively
		\\
		\hline
		$D$& Distance between the center of $\text{Tx}$ and  $\text{Rx}$ at initial time 		
		\\\hline
		$C^{\text{T}}_n$ , $C^{\text{R}}_n$& The first- and last-bounce clusters of the $n$th path, respectively
		\\\hline
		$d_{p,q}^\text{L}(t)$&Distance from ${A}_p^\text{T}$ to  ${A}_q^\text{R}$ at time instant $t$
		\\\hline	
		$d_{m_n}$& The total distance from $A_1^{\text{T}}$ to $A_1^{\text{R}}$  via the $m$th ray of the $n$th cluster at initial time 
		\\\hline
		$d_{p,q,m_n}(t)$&The total distance from $A_p^{\text{T}}$ ($A_q^{\text{R}}$) to $C_n^{\text{T}}$ ($C_n^{\text{R}}$) via the $m$th ray of the $n$th cluster at time instant $t$
		\\\hline
		$\beta^{\text{T}(\text{R})}_{V(H),A}$, $\beta^{\text{T}(\text{R})}_{V(H),E}$&Azimuth and elevation angles of the Tx (Rx) array, respectively
		\\\hline
		$\alpha^{\text{T}(\text{R})}_{A}$,$\alpha^{\text{T}(\text{R})}_{E}$&Azimuth and elevation angles of the mobility of Tx (Rx) array, respectively
		\\\hline
		$\phi^{\text{T}}_{A,L}(t)$, $\phi^{\text{T}}_{E,L}(t)$& Azimuth and elevation angles of departure (AAoD and EAoD)  of the LOS path transmitted from $A_1^{\text{T}}$ at time instant $t$
		\\\hline
		$\phi^{\text{R}}_{A,L}(t)$, $\phi^{\text{R}}_{E,L}(t)$& Azimuth and elevation angles of arrival (AAoA and EAoA) of the LOS path impinging on  $A_1^{\text{R}}$at time instant $t$
		\\\hline
		$\phi^{\text{T}}_{A,m_n}(t)$,$\phi^{\text{T}}_{E,m_n}(t)$&AAoD and EAoD of the $m$th ray in $C^{\text{T}}_n$ transmitted from $A_1^{\text{T}}$ at time instant $t$
		\\\hline
		$\phi^{\text{R}}_{A,m_n}(t)$,$\phi^{\text{R}}_{E,m_n}(t)$&AAoA and EAoA of the $m$th ray in $C^{\text{R}}_n$ impinging on  $A_1^{\text{R}}$ at time instant $t$
		\\\hline
		$v^\text{T},v^\text{R}$&  Velocity of Tx and Rx
		\\\hline
	\end{tabular}}
	
	\label{tab1}
\end{table*}

\subsection{Channel Impulse Response}
Considering small-scale fading, path loss, shadowing, molecular absorption, and blockage effect, the complete channel matrix is given by 
\begin{equation}
{\textbf{H}=[PL\cdot SH\cdot BL\cdot MA]^{\frac{1}{2}}\cdot \textbf{H}_s}
\label{channelmatrix}
\end{equation} 
where $PL$ denotes the path loss. 
{The path loss $PL$ of the channel is modeled by close-in free space reference distance path loss model \cite{PL_model_1,PL_model_2,PL_model_3} as }
{\begin{equation}
PL(d) = PL_0(m)+ 10\gamma \log(\frac{d}{m})
\end{equation}}
{where $PL_0(m)$ is the free space  path loss and can be calculated by Friis equation, $\gamma$  means the propagation coefficient and $m$ is the reference distance.}  
$SH$ denotes the shadowing and is modeled as a lognormal random variable.  The blockage loss $BL$ caused by human activities is taken into account \cite{HuangjieJSAC}. The molecular  absorption loss $MA$ for THz communications can be found in HITRAN database\cite{HITRAN2012}. All these parameters are calculated in power level in this THz channel. 

\newcounter{TempEqCnt} 
\setcounter{TempEqCnt}{\value{equation}} 
\setcounter{equation}{6} 
\begin{figure*}[b]
	\hrulefill  
	\begin{equation}
	\label{rotation1}
	\begin{split}
	\tilde{\phi}^{\text{T}}_{A,L} = \arccos\left(
	\begin{split}
	&\cos(\gamma_y^{\text{T}})\cos(\gamma_z^{\text{T}})\cos({\phi}^{\text{T}}_{A,L})+\\&\left( \sin(\gamma_y^{\text{T}})\cos(\gamma_z^{\text{T}})\cos({\phi}^{\text{T}}_{E,L}-\gamma_x^{\text{T}})-\sin(\gamma_z^{\text{T}})\sin({\phi}^{\text{T}}_{E,L}-\gamma_x^{\text{T}}) \right) \sin({\phi}^{\text{T}}_{A,L})
	\end{split}
	\right)
	\end{split}
	\end{equation}
	\begin{equation}
	\label{rotation2}
	\tilde{\phi}^{\text{T}}_{E,L} = \arg\left(
	\begin{split}
	&\cos(\gamma_y^{\text{T}})\sin({\phi}^{\text{T}}_{A,L})\cos({\phi}^{\text{T}}_{E,L}-\gamma_x^{\text{T}})+\\&
	j
	\left(
	\begin{split}
	&\cos(\gamma_y^{\text{T}})\sin(\gamma_z^{\text{T}})\cos({\phi}^{\text{T}}_{A,L})+
	\\
	&\left( \sin(\gamma_y^{\text{T}})\sin(\gamma_z^{\text{T}})\cos({\phi}^{\text{T}}_{E,L}-\gamma_x^{\text{T}})-\cos(\gamma_z^{\text{T}})\sin({\phi}^{\text{T}}_{E,L}-\gamma_x^{\text{T}}) \right) \sin({\phi}^{\text{T}}_{A,L})
	\end{split}
	\right) 
	\end{split}
	\right) .
	\end{equation}
\end{figure*}
\setcounter{equation}{\value{TempEqCnt}} 

Due to the large bandwidth of THz communication systems, it is not suitable to set the whole channel parameters at center frequency. {To describe the channel more accurately, the communication band $\text{\textbf{B}}$ is divided into $N_F$ small sub-bands with bandwidth of $\textbf{B}_\text{sub}=\textbf{B}/N_F$. The center frequency of the $i$th sub-band is $f_i$. The channel matrix of the $i$th sub-band is denoted as $\textbf{H}_{f_i}$. Considering the nonnegligible difference in different frequency sub-bands, both large-scale parameters and small-scale fading parameters are calculated for different sub-bands separately. The large-scale fading parameters in (\ref{channelmatrix}), \textit{i.e.}, $PL$, $SH$, $BL$, and $MA$ are updated for different frequency sub-bands. The small-scale channel matrix of a sub-band is denoted as $\textbf{H}_{s,f_i}$.  In traditional communication systems, the received signal can be calculated by the convolution of the transmitted signal and the channel impulse response in the time domain or their product in the frequency domain. However, in the communication systems with large bandwidth, the transmitted signal can be also divided into the sub-band signals. After each sub-band experiencing its corresponding channel matrix, the received signal can be calculated by summing up all the band-pass signals as  }

{
\begin{equation}
R(t,\tau)=\sum_{i=0}^{N_F}\int_{-\infty}^{\infty}S(t,f)\cdot A_{f_i}(f)\cdot H_{f_i}(t,f)\cdot w(t)+n(t)
\end{equation}
}
{
where $S(t,f)$ is the spectrum of the transmitted signal, and $R(t,\tau)$ is the received signal. The filtering function $A_{f_i}(f)$ denotes the ideal bandwidth band-pass filter for the sub-band. The transfer function matrix $H_{f_i}(t,f)$ is the Fourier transform of $\textbf{H}_{f_i}$, $w(t)$ is the time-variant beamforming matrix, }
{and $n(t)$ is the normalized complex additive white Gaussian noise. }
The channel impulse response of the $i$th sub-band $\textbf{H}_{s,f_i}=[h_{p,q,f_i}(t,\tau)]_{M^\text{R}\times M^\text{T}}$ where $M^\text{R}$ and $M^\text{T}$ are the element numbers of Tx and Rx, respectively. $h_{p,q,f_i}(t,\tau)$ is the CIR from $A_p^{\text{T}}$ to $A_q^{\text{R}}$ and expressed as the summation of the line-of-sight (LOS) and non-LOS (NLOS) components, i.e.,
\begin{equation}
\label{CIR_total}
\begin{split}
h_{p,q,f_i}(t,\tau)=&\sqrt{\frac{K^\text{R}}{K^\text{R}+1}}h^\text{L}_{p,q,f_i}(t,\tau)
\\
&+\sqrt{\frac{1}{K^\text{R}+1}}\sum_{n=1}^{N_{p,q}(t)}\sum_{m=1}^{M_{n,f_i}}h^\text{N}_{p,q,f_i,m_n}(t,\tau)
\end{split}
\end{equation}
where $K^\text{R}$ is the K-factor, $N_{p,q}(t)$ is the number of clusters from $A_p^{\text{T}}$ to $A_q^{\text{R}}$ at time instant $t$. $M_{n,f_i}$ represents the  time-space variant number of rays in the $n$th cluster of the $i$th sub-band. It  is frequency dependent and assumed to follow a Poisson distribution Pois($\tilde{\lambda}$) \cite{J_JSAC14_Akdeniz} where $\tilde{\lambda}$ is the mean and variance of $M_{n,f_i}$. 

For the LOS components in (\ref{CIR_total}), the complex channel gain can be expressed as
{
\begin{equation}
\label{CIR_LOS}
\begin{split}
&h^\text{L}_{p,q,f_i}(t,\tau) =
\\& \left[ 
\begin{matrix}
F_{q,V}(\tilde{\phi}^{\text{R}}_{E,L}(t),\tilde{\phi}^{\text{R}}_{A,L}(t))
\\
F_{q,H}(\tilde{\phi}^{\text{R}}_{E,L}(t),\tilde{\phi}^{\text{R}}_{A,L}(t))
\end{matrix}
\right]^{T}
\left[ 
\begin{matrix}
e^{j\theta_{LOS}} & 0
\\
0&-e^{j\theta_{LOS}}  
\end{matrix}
\right]
\\&
\left[ 
\begin{matrix}
F_{p,V}(\tilde{\phi}^{\text{T}}_{E,L}(t),\tilde{\phi}^{\text{T}}_{A,L}(t))
\\
F_{p,H}(\tilde{\phi}^{\text{T}}_{E,L}(t),\tilde{\phi}^{\text{T}}_{A,L}(t))
\end{matrix}
\right] e^{j2\pi\nu_{p,q,f_i}^{LOS}(t)t}\delta(\tau-\tau_{p,q}^\text{L}(t))
\end{split}
\end{equation}
where $\{\cdot\}^{T}$ denotes transposition operation}, $\theta_{LOS}$ is uniformly distributed within $(0,2\pi]$, $F_{p(q),V}$ and $F_{p(q),H}$ are the antenna patterns of $A_p^{\text{T}}$ and $A_q^{\text{R}}$ at vertical and horizontal planes, respectively. Note that in this paper, $\tilde{\phi}$ represents angles in the LCS and needs to be transformed from angles in the GCS. {The angle in LCS is obtained by  rotating from the corresponding angles in the GCS for  three axes \cite{3GPP38901}. Similar to chapter 7.1 in \cite{3GPP38901}, rotation of  $x_\text{G}$ / $y_\text{G}$ / $z_\text{G}$ is denoted by $\gamma_x^{\text{T}}$ / $\gamma_y^{\text{T}}$ / $\gamma_z^{\text{T}}$, respectively. For example, $\tilde{\phi}^{\text{T}}_{A,L}$ and $\tilde{\phi}^{\text{T}}_{E,L}$ can be obtained by (\ref{rotation1}) and (\ref{rotation2}), respectively}

Then the antenna patterns in (\ref{CIR_LOS}) can be calculated as 
\setcounter{equation}{8}
\begin{equation}
F_{q,V}(\tilde{\phi}_{E},\tilde{\phi}_{A}) = G(\tilde{\phi}_{E},\tilde{\phi}_{A})\sin(\tilde{\phi}_{E})
\end{equation}
\begin{equation}
F_{q,H}(\tilde{\phi}_{E},\tilde{\phi}_{A}) = G(\tilde{\phi}_{E},\tilde{\phi}_{A})\cos(\tilde{\phi}_{E}).
\end{equation}
{In the proposed channel model, antenna arrays at Tx and Rx sides can be any radiation pattern for each element. The radiation pattern can be substituted into (6) to generate the CIR of the LOS path. In this paper, we assume that the dipole antennas are utilized at both Tx and Rx sides.} In this case \cite{antennapattern}, we have
\begin{equation}
G(\tilde{\phi}_{E},\tilde{\phi}_{A}) = \sqrt{1.64}\frac{\cos(\frac{\pi}{2}\cos\tilde{\phi}_{A})}{\sin\tilde{\phi}_{A}}.
\end{equation}

The delay $\tau_{p,q}^\text{L}(t)$ is calculated by $\tau_{p,q}^\text{L}(t) ={d}_{p,q}^\text{L}(t)/c $, where $c$ is the speed of light
and ${d}_{p,q}^\text{L}(t)$ denotes the distance from ${A}_p^\text{T}$ to  ${A}_q^\text{R}$ at time instant $t$. The vector $\overrightarrow{d}_{p,q}^\text{L}(t)$ can be calculated as
\begin{equation}
\overrightarrow{d}_{p,q}^\text{L}(t) = \overrightarrow{D}+\overrightarrow{A}_p^{\text{R}}-\overrightarrow{A}_q^{\text{T}}+\int_{0}^\text{T}(\overrightarrow{v}^\text{R}(t)-\overrightarrow{v}^\text{T}(t))dt.
\end{equation}
In this model, we assume that the velocity vector of Tx or Rx is constant and can be 
expressed as
\begin{equation}
\overrightarrow{v}^\text{T}={v}^\text{T}\cdot[\cos\alpha^{\text{T}}_{E}\cos\alpha^{\text{T}}_{A}, \cos\alpha^{\text{T}}_{E}\sin\alpha^{\text{T}}_{A}, \sin\alpha^{\text{T}}_{E}]
\end{equation}
\begin{equation}
\overrightarrow{v}^\text{R}={v}^\text{R}\cdot[\cos\alpha^{\text{R}}_{E}\cos\alpha^{\text{R}}_{A}, \cos\alpha^{\text{R}}_{E}\sin\alpha^{\text{R}}_{A}, \sin\alpha^{\text{R}}_{E}].
\end{equation}

The Doppler frequency $\nu_{p,q,f_i}^\text{L}(t)$ between $A_p^{\text{T}}$ and $A_q^{\text{R}}$ is expressed as
\begin{equation}
\nu_{p,q,f_i}^\text{L}(t) = \frac{1}{\lambda(f_i)}\frac{\left\langle \overrightarrow{d}_{p,q}^\text{L}(t), \overrightarrow{v}^{\text{R}}-\overrightarrow{v}^{\text{T}} \right\rangle }{\|\overrightarrow{d}_{p,q}^\text{L}(t)\|}
\end{equation}
where $\left\langle\cdot \right\rangle $ is the inner product operator and $\|\cdot\|$ calculates the Frobenius norm. The 
$\lambda(f_i)$  is the wavelength of the $i$th sub-band and calculated as 
\begin{equation}
\lambda(f_i) = \frac{c}{f_i}.
\end{equation}

For NLOS components, the channel gain $h^\text{N}_{p,q,f_i,m_n}(t)$ can be written as
{
\begin{equation}
\begin{split}
&h^\text{N}_{p,q,f_i,m_n}(t,\tau)=
\left[ 
\begin{matrix}
F_{q,V}(\tilde{\phi}^{\text{R}}_{E,m_n}(t),\tilde{\phi}^{\text{R}}_{A,m_n}(t))
\\
F_{q,H}(\tilde{\phi}^{\text{R}}_{E,m_n}(t),\tilde{\phi}^{\text{R}}_{A,m_n}(t))
\end{matrix}
\right]^{T}
\\
&
\left[ 
\begin{matrix}
e^{j\theta^{VV}_{m_n}} & \sqrt{\kappa^{-1}_{m_n}}e^{j\theta^{VH}_{m_n}}
\\
\sqrt{\kappa^{-1}_{m_n}}e^{j\theta^{HV}_{m_n}}&e^{j\theta^{HH}_{m_n}} 
\end{matrix}
\right]\left[ 
\begin{matrix}
F_{p,V}(\tilde{\phi}^{\text{T}}_{E,m_n}(t),\tilde{\phi}^{\text{T}}_{A,m_n}(t))
\\
F_{p,H}(\tilde{\phi}^{\text{T}}_{E,m_n}(t),\tilde{\phi}^{\text{T}}_{A,m_n}(t))
\end{matrix}
\right]
\\
& 
\sqrt{P_{p,q,f_i,m_n}(t)}e^{j2\pi \nu_{p,q,m_n}^{\text{T}}(t)}e^{j2\pi \nu_{p,q,m_n}^{\text{R}}(t)}\delta(\tau-\tau_{p,q,m_n}(t))
\end{split}
\end{equation}
}
where $\kappa_{m_n}$ stands for the cross polarization power ratio, $\theta^{VV}_{m_n}$, $\theta^{VH}_{m_n}$, $\theta^{HV}_{m_n}$, and $\theta^{HH}_{m_n}$ are initial phases with uniform distribution over $(0,2\pi]$. $P_{p,q,f_i,m_n}$ is the powers of the $m$th ray in the $n$th cluster between $A_p^{\text{T}}$ and $A_q^{\text{R}}$ at time instant $t$. 
The delay of the $m$th in the $n$th cluster is denoted as  $\tau_{p,q,m_n}(t)$ and calculated as $\tau_{p,q,m_n}(t) = d_{p,q,m_n}(t)/c$. The Doppler frequencies at Tx and Rx side are denoted as  $\nu_{p,q,m_n}^{\text{T}}(t)$ and $\nu_{p,q,m_n}^{\text{R}}(t)$, respectively. {The total number of paths in each cluster can be set as a typical number such as 50 or 100 in the channel simulation.}

In order to generate the complete THz channel coefficient, we need firstly to generate a set of clusters at initial time and first element of Tx/Rx in the first band. Then the STF cluster  evolution will be taken in the space-time-frequency domains and parameters will also be updated. In the same time,  the new clusters have the probability of birth process. 

For a new cluster, spherical wavefront needs to be considered due to the utilization of large antenna arrays. The time variant transmission distances caused by long traveling movement are also nonnegligible. {In THz communication systems, the specular reflection accounts the main part in the received power and diffuse scattering is only detectable around the specular reflection points\cite{RN197}. For a specular reflection path, the position of reflection point will move when the Tx or Rx moves. Different from other GBSMs in \cite{massiveMIMO1} where the total distance is divided into three parts: from the Tx to the first cluster, form the Rx to the last cluster, and the virtual link between these two clusters. In this channel model, we calculate the total distance of the path directly, the total distance from the Tx to the Rx is considered as an integrated random variable. The total distance of cluster is described recursively with an interarrival distance relative to the previous cluster } 
{
\begin{equation}
d_n = d_{n-1}+\Delta d_n
\end{equation}
}
{where $\Delta d_n$ is an negative exponential distribution (NEXP) random variable\cite{RN194} with parameter $\bar{d}^\text{N}$. For the first cluster, $\Delta d_1$ means the distance difference relative to the LoS path.}
{For the first cluster, $\Delta d_1$ means the distance difference relative to the direct distance between the Tx and Rx, which is the LOS path in the LOS scenario and is still valid in the NLOS scenario..}

{The angular parameters of each ray consist of two parts: the angle of the cluster center and the relative angle. The relative angles $\Delta \phi^{\text{T}}_{A,m_n}$, $\Delta \phi^{\text{T}}_{E,m_n}$, $\Delta \phi^{\text{R}}_{A,m_n}$, and $\Delta \phi^{\text{T}}_{R,m_n}$ calculate the angular offset compared to the center of clusters. 
The angles can be calculated as 
\begin{equation}
\begin{split}
&[\phi^{\text{T}}_{A,m_n},  \phi^{\text{T}}_{E,m_n},  \phi^{\text{R}}_{A,m_n},  \phi^{\text{R}}_{E,m_n}] =
[\phi^{\text{T}}_{A,n}, \phi^{\text{T}}_{E,n}, \phi^{\text{R}}_{A,n}, \phi^{\text{R}}_{E,n}]
\\&
+[\Delta \phi^{\text{T}}_{A,m_n}, \Delta \phi^{\text{T}}_{E,m_n}, \Delta \phi^{\text{R}}_{A,m_n}, \Delta \phi^{\text{R}}_{E,m_n}].
\end{split}
\end{equation}}

The angular parameters $\phi^{\text{R}}_{A,n}$, $\phi^{\text{R}}_{E,n}$ of cluster are assumed to obey wrapped Gaussian distributions. Angles of arrival (AoAs) of the $n$th cluster are generated as
\begin{equation}
\phi^{\text{R}}_{E,n} = \text{std}[\phi^{\text{R}}_{E,n}] Y_{E,n}+\psi_{E,n}
\end{equation}
\begin{equation}
\phi^{\text{R}}_{A,n} = \text{std}[\phi^{\text{R}}_{A,n}] Y_{A,n}+\psi_{A,n}
\end{equation}
where $Y_{E,n}$ and $Y_{A,n}$ follow the Gaussian distribution $N(0,1)$, $\text{std}[\phi^{\text{R}}_{E,n}]$ and $\text{std}[\phi^{\text{R}}_{A,n}]$ are standard deviations of AoAs and need to be estimated.

In this model, we assume that all the diffusely scattered rays happen around the specular reflected path according to \cite{RN369}. When modeling the relative angle, we assume that the center of the cluster is the specular point. All the scatterers around the  specular point is Gaussian distributed. The  standard variances of relative angle  $\Delta \phi^{\text{T}}_{A,m_n}$, $\Delta \phi^{\text{T}}_{E,m_n}$, $\Delta \phi^{\text{R}}_{A,m_n}$, and $\Delta \phi^{\text{T}}_{R,m_n}$ are denoted as $\sigma^{\text{T}}_{A,n}$, $\sigma^{\text{T}}_{E,n}$, $\sigma^{\text{R}}_{A,n}$, and $\sigma^{\text{R}}_{E,n}$, respectively.

The total distance of rays in each cluster $d_{p,q,m_n}$  can be calculated from the relative angles  and distances of specular reflection $d_{p,q,n}$ by the  geometric  relationship. 

\begin{figure}[tb]
	\centerline{\includegraphics[width=0.5\textwidth]{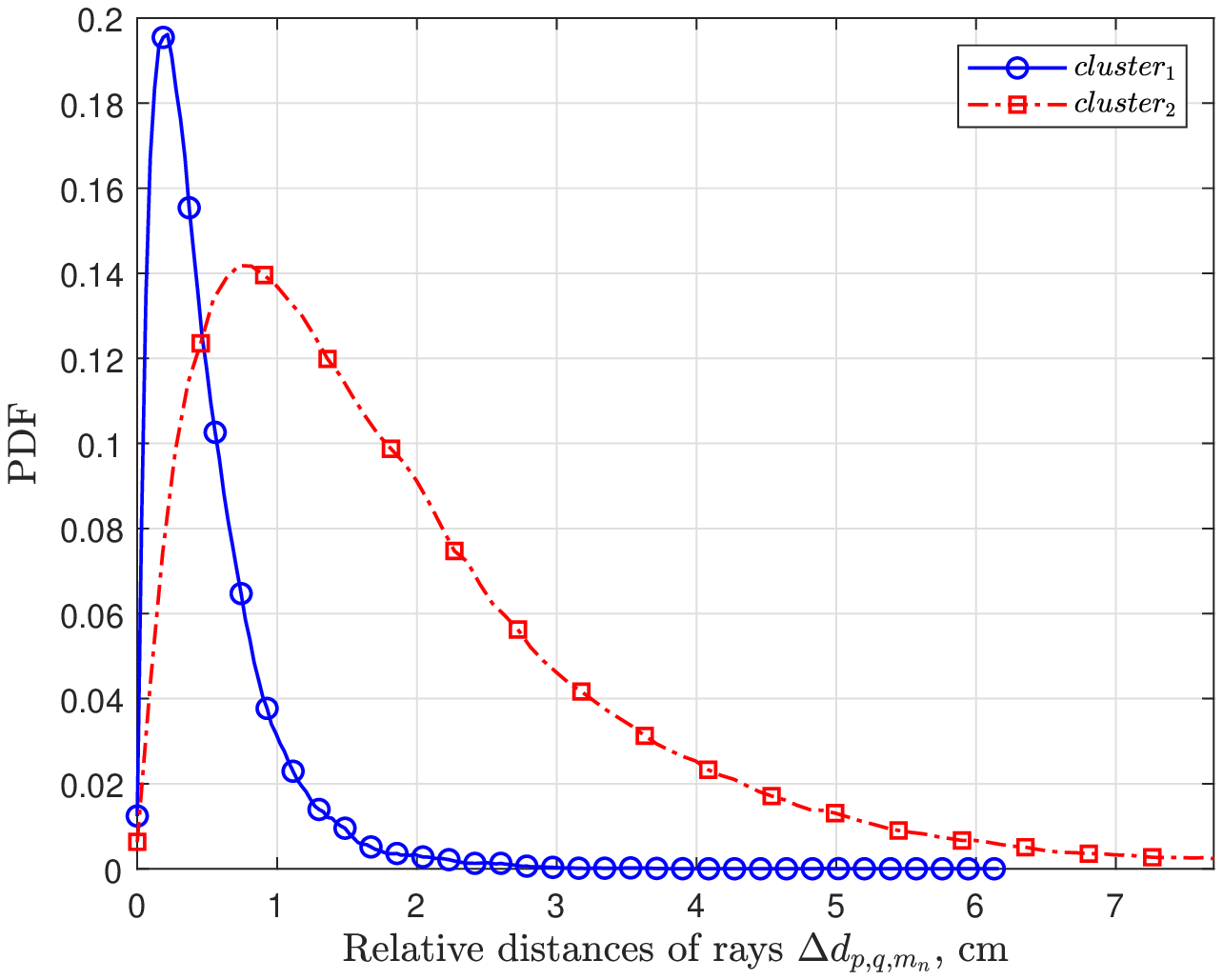}}
	\caption{{PDF  of the relative distances within clusters ($d_{p,q,1}$ = 8 m, $r^{\text{T}}_{c}$ = 0.4, $r^{\text{R}}_{c}$ = 0.6, $\sigma^{\text{T}}_{A,1}$ = $\sigma^{\text{T}}_{E,1}$ =  $\sigma^{\text{R}}_{A,1}$ = $\sigma^{\text{R}}_{E,1} = 2^\circ$, $\sigma^{\text{T}}_{A,2}$ = $\sigma^{\text{T}}_{E,2}$ =  $\sigma^{\text{R}}_{A,2}$ = $\sigma^{\text{R}}_{E,2} = 4^\circ$).}}
	\label{pdf}
\end{figure}

According to the relative angle, the total distance of different scattering paths can be calculated 
\begin{equation}
d_{p,q,m_n}=\sqrt{{d^{\text{V}}_{p,q,m_n}}^2+{d^{\text{H}}_{p,q,m_n}}^2}
\end{equation}
where $d^{\text{V}}_{p,q,m_n}$ and $d^{\text{H}}_{p,q,m_n}$ represent the vertical and  horizontal distances of the path length, respectively. They can be calculated by
{
\begin{equation}
\begin{split}
{d^{\text{V}}_{p,q,m_n}}& = {d_{p,q,n}} \sin(\phi^{\text{R}}_{A,n})r^{\text{R}}_{c}/\cos(\Delta\phi_{E,m_n}^{\text{R}})
\\&
+{d_{p,q,n}} \sin(\phi^{\text{T}}_{A,n})r^{\text{T}}_{c}/\cos(\Delta\phi_{E,m_n}^{\text{T}})
\end{split}
\end{equation}
\begin{equation}
\begin{split}
{d^{\text{H}}_{p,q,m_n}}& = {d_{p,q,n}} \cos(\phi^{\text{R}}_{A,n})r^{\text{R}}_{c}/\cos(\Delta\phi_{A,m_n}^{\text{R}})
\\&+{d_{p,q,n}} \cos(\phi^{\text{T}}_{A,n})r^{\text{T}}_{c}/\cos(\Delta\phi_{A,m_n}^{\text{T}})
\end{split}
\end{equation}
}
where $r^{\text{T(R)}}_{c}$ is the ratio of the distance between the cluster and the Tx(Rx) to the  total distance. {When the relative angles are zero, the path can be considered as specular reflection. Then, the relative distance can be obtained by $d_{p,q,m_n}$ and the specular path.}For single bounce clusters, it is clear that $r^{\text{T}}_{c}$+$r^{\text{R}}_{c}$ = 1. For multiple bounce rays, we have $r^{\text{T}}_{c}$+$r^{\text{R}}_{c}\textless1$. 
The simulated probability density functions (PDF) for different variances are shown in Fig. \ref{pdf}. 

\subsection{STF Evolution}
\begin{figure}[tb]
	\centerline{\includegraphics[width=0.5\textwidth]{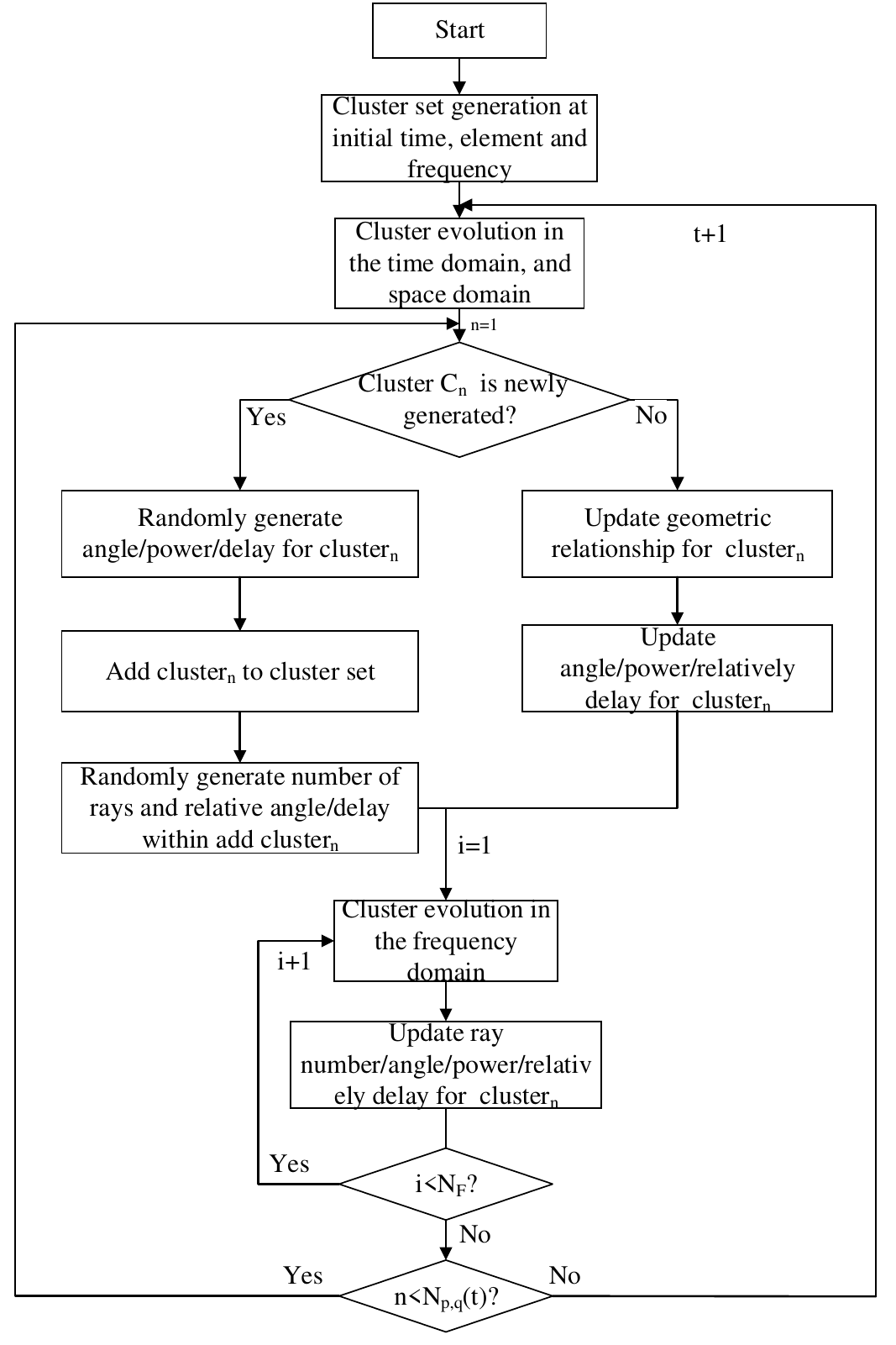}}
	\caption{Flowchart of the STF cluster evolution for the proposed THz 3D GBSM.}
	\label{fig_2}
\end{figure}

In this sub-section, we will generate the STF-variant parameters for the channel. The evolution in time axis and array axis is caused by the offset of geometry relationship. The evolution in the frequency domain is caused by frequency-dependent scattering.  

birth-death process has been validated by channel measurements in vehicle-to-vehicle scenarios \cite{Bianji1}. In massive MIMO communication
systems, similar properties can be observed on the array axis.
In this paper, space-time cluster evolutions are modeled in a uniform manner.
The flowchart of the STF cluster evolution based on the birth-death process is shown in Fig.~\ref{fig_2} .  
The generation (birth) and recombination (death) rates of clusters are denoted as {$\lambda_G$ and $\lambda_\text{R}$}, respectively. Firstly,  initial parameters for clusters are generated at time $t$ and first element of Tx and Rx. Then, the unified array-time cluster evolution are conducted and parameters at $t+\Delta t$ and other elements are updated according to the geometric relationship. 

For the Tx side, the probability of a cluster remains over the time interval $\Delta t$ and antenna element spacing $\tilde{\delta}_{p}$ can be calculated as
\begin{equation}
\begin{split}
&P_{\text {remain }}^\text{T}\left(\Delta t, \tilde{\delta}_{p}\right)
\\&
=\exp \left(-\lambda_\text{R}\left[\left(\tilde{\varepsilon}_{1}^\text{T}\right)^{2}+\left(\varepsilon_{2}^\text{T}\right)^{2}-2 \tilde{\varepsilon}_{1}^\text{T} \varepsilon_{2}^\text{T} \cos \left(\alpha_{A}^\text{T}-\tilde{\beta}_{A}^\text{T}\right)\right]^{\frac{1}{2}}\right)
\end{split}
\end{equation}
where 
\begin{equation}
\tilde{\varepsilon}_{1}^\text{T}=\tilde{\delta}_{p} \cos \left(\tilde{\beta}_{E}^\text{T}\right) / D_{c}^{A}
\end{equation}
\begin{equation}
\varepsilon_{2}^\text{T}=v^\text{T} \Delta t / D_{c}^{S}.
\end{equation}
Different from the evolution along the uniform linear array (ULA) in \cite{RN322,massiveMIMO1} the evolution along the UPA is not in one fixed direction. For the elements in the first row and first column, the $\tilde{\delta}_{p}$ is ${\delta}_{p}$, the $\tilde{\beta}_{E}^\text{T}$ equals to ${\beta}_{H,E}^\text{T}$ and ${\beta}_{V,E}^\text{T}$, respectively. 
For other elements in the array, we have $\tilde{\delta}_{p} = \sqrt{2}{\delta}_{p}$ and $\tilde{\beta}_{E}^\text{T} = ({\beta}_{H,E}^\text{T}+{\beta}_{V,E}^\text{T})/2$. Symbols $D_{c}^{A}$ and $D_{c}^{S}$ are scenario-dependent correlation factors in the
array and time domains, respectively.

For the Rx side, the probability of a cluster existing over the time interval $\Delta t$ and element spacing $\tilde{\delta}_q$ is calculated similarly. Since each antenna element has its own observable cluster set, only a cluster can be seen by at least one Tx antenna and one Rx antenna, it can contribute to the received power. Therefore, the joint probability of a cluster existing over $\Delta t$ and element spacing $\tilde{\delta}_p, \tilde{\delta}_q$ is calculated as
\begin{equation}
P_{\text {remain }}\left(\Delta t, \tilde{\delta}_{p}, \tilde{\delta}_{q}\right)=P_{\text {remain }}^\text{T}\left(\Delta t, \tilde{\delta}_{p}\right) \cdot P_{\text {remain }}^\text{R}\left(\Delta t, \tilde{\delta}_{q}\right).
\end{equation}
The mean number of newly generated clusters is obtained by
\begin{equation}
\mathbb{E}\left\{N_{\text {new }}\right\}=\frac{\lambda_{G}}{\lambda_{R}}\left[1-P_{\text {remain }}\left(\Delta t, \tilde{\delta}_{p}, \tilde{\delta}_{q}\right)\right].
\end{equation}

\begin{figure*}[tb]
	\begin{minipage}[t]{0.5\linewidth}
		\centerline{\includegraphics[width=1\textwidth]{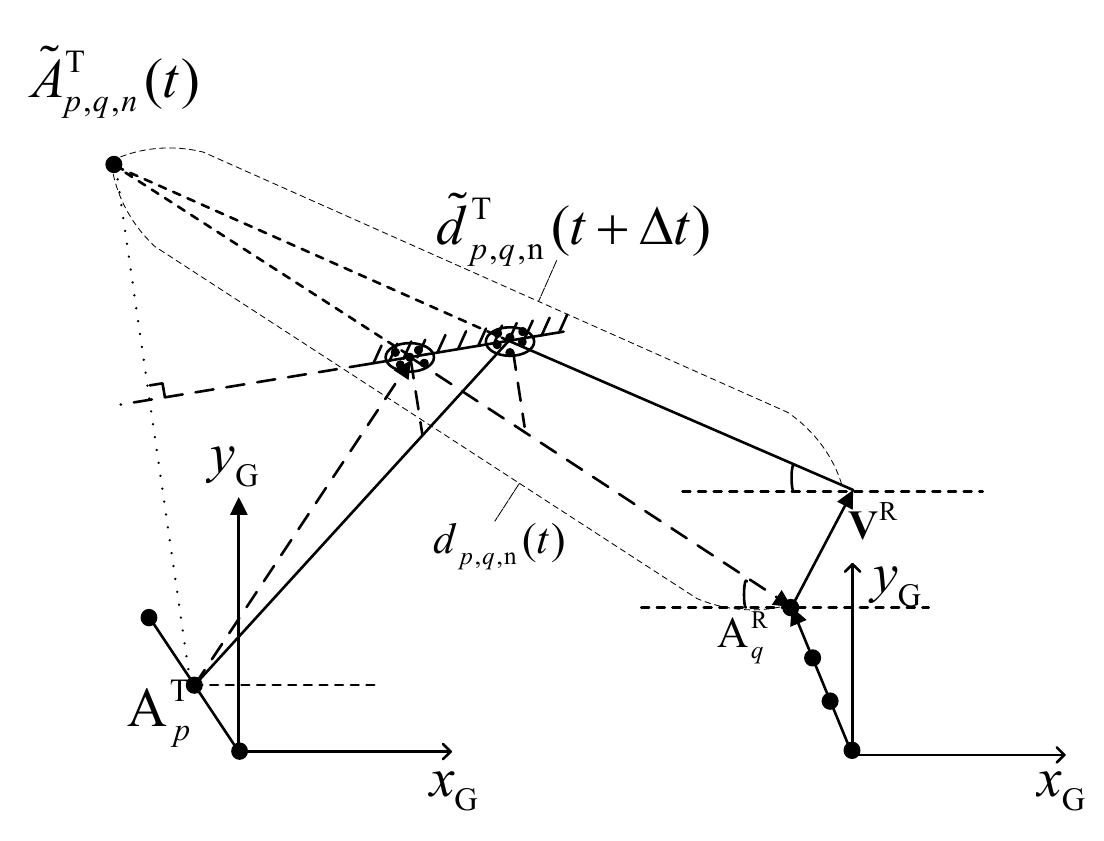}}
		\caption*{(a)}
		\label{fig_3a}
	\end{minipage}
	\begin{minipage}[t]{0.5\linewidth}
	\centerline{\includegraphics[width=1\textwidth]{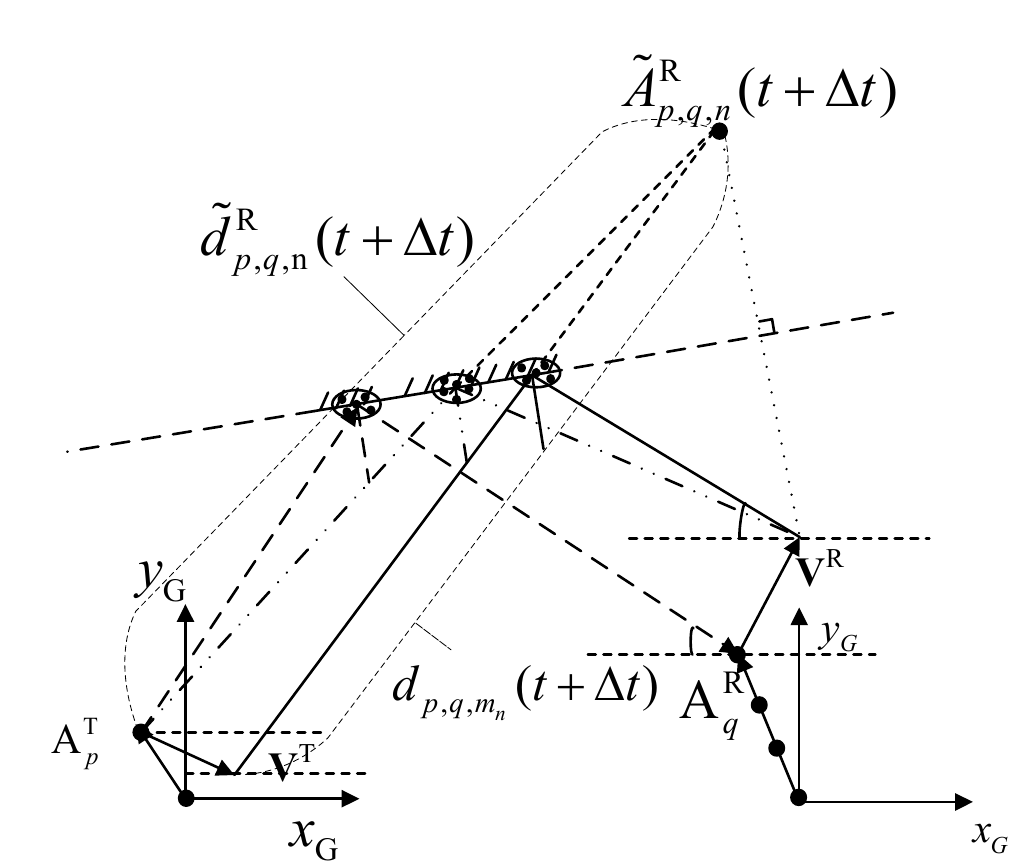}}
	\caption*{(b)}
	\label{fig_3b}
\end{minipage}
\begin{minipage}[t]{0.5\linewidth}
	\centerline{\includegraphics[width=1\textwidth]{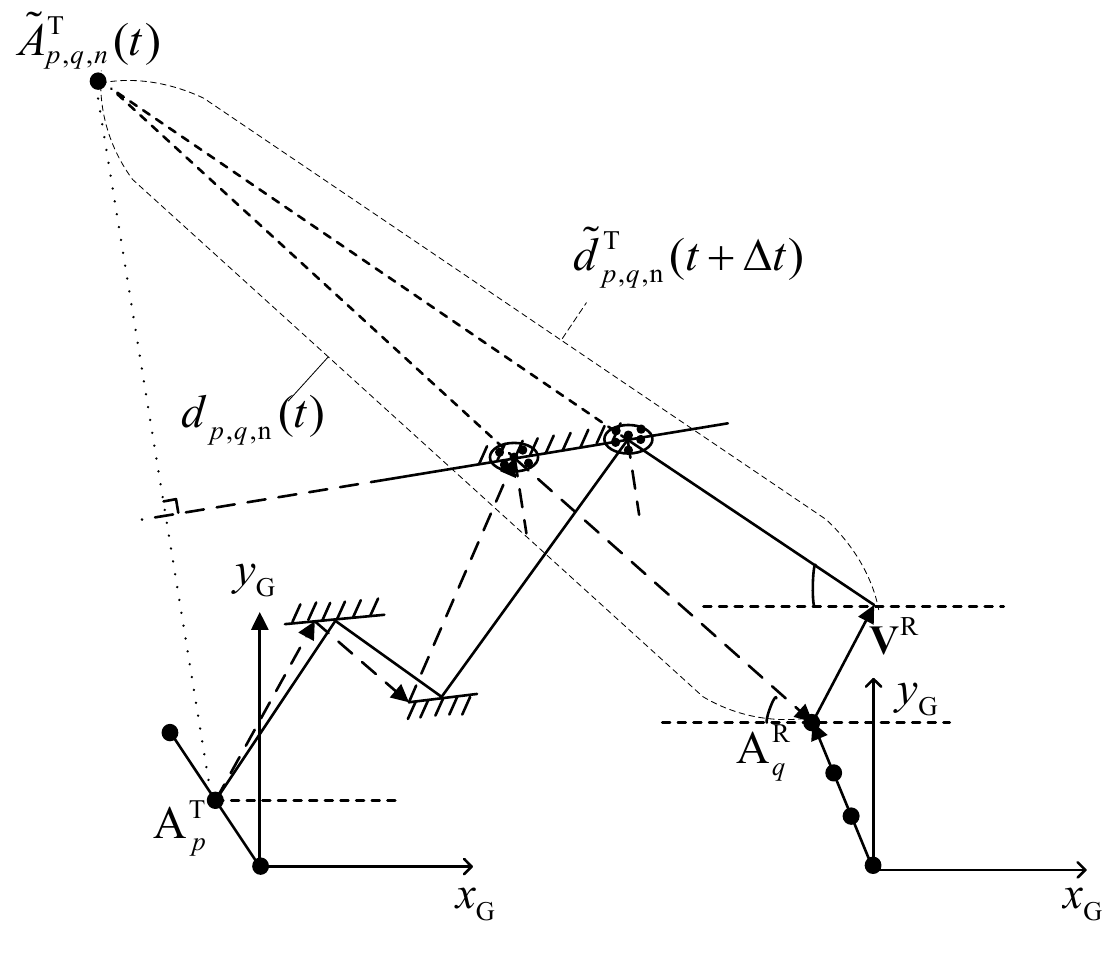}}
	\caption*{(c)}
	\label{fig_3c}
\end{minipage}
\begin{minipage}[t]{0.5\linewidth}
	\centerline{\includegraphics[width=1\textwidth]{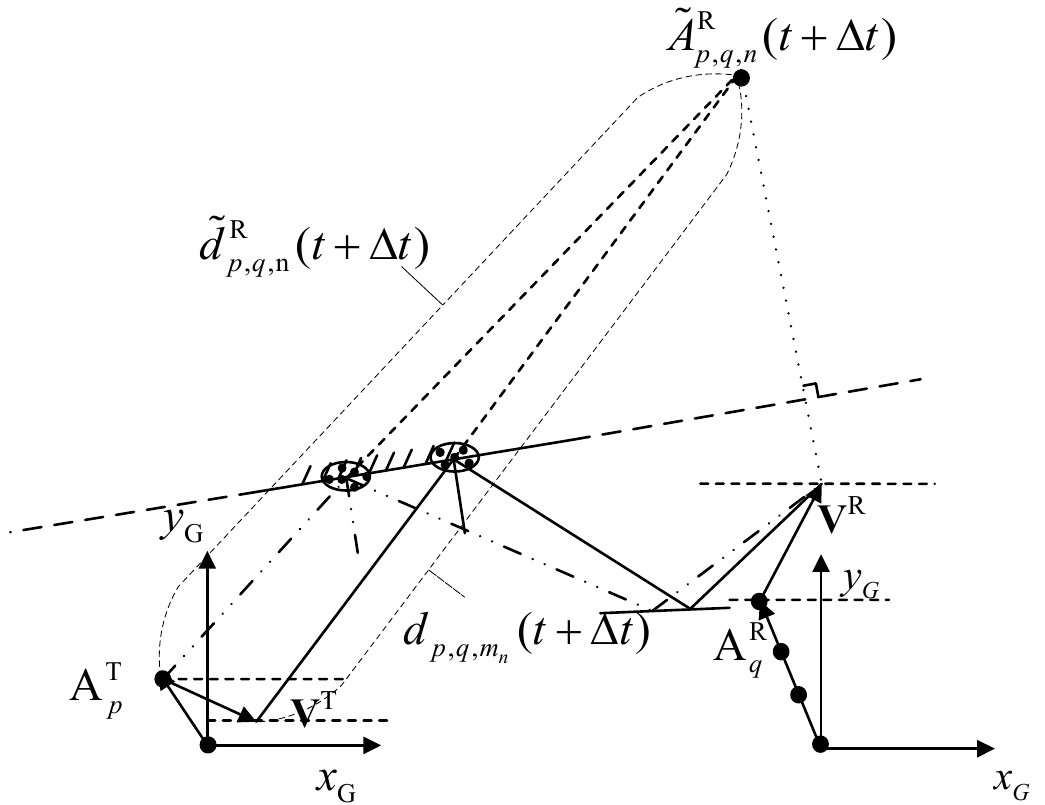}}
	\caption*{(d)}
	\label{fig_3d}
\end{minipage}
\caption{ {Array-time evolution for (a) single bounce path in the Rx side, (b) single bounce path in the Tx side, (c) multi bounce path in the Rx side, and (d) mulit bounce path in the Tx side} }
\label{fig_4}
\end{figure*}

Since the center of all of the clusters are specular reflection, there exists a mirror point for $A^\text{T}_p$ and $A^\text{T}_p$.  As we can see in the Fig.~\ref{fig_4}, the mirror point of $A^\text{T}_p$ is symmetrical about the first reflection surface of the $n$th path at time instant $t$ and denoted as $\tilde{A}^\text{T}_{p,q,n}(t)$. Similarly, the mirror point of $A^\text{R}_q$ is symmetry about the first reflection surface of the $n$th path at time instant $t$ and denoted as $\tilde{A}^\text{R}_{p,q,n}(t)$.

According to the geometrical relationships in Fig. 1, the distance vectors of the $m$th ray of the $n$th cluster $\overrightarrow{d}_{p,q,m_n}^{\text{T}}(t)$ and $\overrightarrow{d}_{p,q,m_n}^{\text{R}}(t)$ at the Tx side and Rx side are calculated as
\begin{equation}
\overrightarrow{d}_{p,q,m_n}^{\text{T}}(t) = d_{p,q,m_n}^{\text{T}}(t)
\left[
\begin{matrix}
\cos\phi^{\text{T}}_{E,m_n}\cos\phi^{\text{T}}_{A,m_n}
\\
\cos\phi^{\text{T}}_{E,m_n}\sin\phi^{\text{T}}_{A,m_n}
\\
\sin\phi^{\text{T}}_{E,m_n}
\end{matrix}
\right] 
\end{equation}
\begin{equation}
\overrightarrow{d}_{p,q,m_n}^{\text{R}}(t) = d_{p,q,m_n}^{\text{R}}(t)
\left[
\begin{matrix}
\cos\phi^{\text{R}}_{E,m_n}\cos\phi^{\text{R}}_{A,m_n}
\\
\cos\phi^{\text{R}}_{E,m_n}\sin\phi^{\text{R}}_{A,m_n}
\\
\sin\phi^{\text{R}}_{E,m_n}
\end{matrix}
\right].
\end{equation}
It is clear that the total distance of the path equals to the distance from the mirror point to the antenna in another side. We have that 
\begin{equation}
{d}_{p,q,m_n}(t) = \|\overrightarrow{d}_{p,q,m_n}^{\text{T}}(t)\|= \|\overrightarrow{d}_{p,q,m_n}^{\text{R}}(t)\|.
\end{equation}
For the array time evolution, the Tx and Rx are calculated separately. In the first step, we assume that Rx  moving along time-array axis and Tx is static. 
We have that 
\begin{equation}
\begin{split}
\tilde{d}_{p,q,m_n}(t+\Delta t) &= \left\| \overrightarrow{\tilde{d}}_{p,q,m_n}^\text{R}(t+\Delta t)\right\| \\&=\left\| \overrightarrow{\tilde{d}}_{1,1,m_n}^\text{R}(t)+\overrightarrow{A}^\text{R}_q+\overrightarrow{v}^\text{R}\cdot\Delta t\right\|
\end{split}
\end{equation}
where $\tilde{d}_{p,q,m_n}(t+\Delta t)$ is a temporary distance after the first step. The $\overrightarrow{\tilde{d}}_{p,q,m_n}^\text{R}(t+\Delta t)$ is the vector from the mirror point of Tx to the ${A}^\text{R}_q$. 
Then the Tx is moving and Rx keep static. The new distance  can be calculated as 
\begin{equation}
\begin{split}
{d}_{p,q,m_n}(t+\Delta t) &= \left\|\overrightarrow{d}_{p,q,m_n}(t+\Delta t)\right\| \\&= \left\|\overrightarrow{\tilde{d}}_{p,q,m_n}^\text{T}(t+\Delta t)-\overrightarrow{A}^\text{T}_p-\overrightarrow{v}^\text{T}\cdot\Delta t\right\|
\end{split}
\end{equation}
where $\overrightarrow{\tilde{d}}_{p,q,m_n}^\text{T}(t+\Delta t)$ denotes the vector from the mirror point of Rx to the ${A}^\text{T}_p$.
The angle can also be calculated from the updated vector. {For both single-bounce paths and multi-bounce paths, we only consider the total path lengths which contain the virtual link in multi-bounce paths.}

The Doppler frequency $\nu_{p,q,m_n}^{\text{T}}(t)$ at Tx and $\nu_{p,q,m_n}^{\text{R}}(t)$ are calculated as 
\begin{equation}
\nu_{p,q,m_n}^{\text{T}}(t)= \frac{1}{\lambda(f_i)}\frac{\left\langle \overrightarrow{d}_{p,q,m_n}^\text{T}(t), \overrightarrow{v}^{\text{T}} \right\rangle }{\|\overrightarrow{d}_{p,q,m_n}^\text{T}(t)\|}
\end{equation}
\begin{equation}
\nu_{p,q,m_n}^{\text{R}}(t)= \frac{1}{\lambda(f_i)}\frac{\left\langle \overrightarrow{d}_{p,q,m_n}^\text{R}(t), \overrightarrow{v}^{\text{R}} \right\rangle }{\|\overrightarrow{d}_{p,q,m_n}^\text{R}(t)\|}.
\end{equation}

For different sub-bands, positions of cluster and Tx/Rx remain unchanged so that the birth-death process of clusters are not considered. However, considering the scattering in different sub-bands, the birth-death process for rays in each clusters are considered along the frequency axis. The probability of a ray remains over two adjacent sub-channel is calculated as
\begin{equation}
P_{\text {remain }}\left(B_\text{sub}\right) = \text{exp}\left( {\frac{-\lambda_\text{R} \cdot\rho_\text{S}\cdot B_\text{sub}}{B^f_c}}\right) 
\end{equation}
where $\rho_\text{S}$ is a coefficient related to the reflection surface. Symbol $B^f_c$ is scenario-dependent correlation factors in the frequency domain.
Then the mean number of newly generated rays is obtained by
\begin{equation}
\mathbb{E}\left\{N_{\text{new}}\right\}=\frac{\lambda_{G}}{\lambda_{R}}[1-P_{\text {remain}}\left(B_\text{sub}\right)].
\end{equation}

For the survived rays, the space-time-frequency varying ray power between $A^{\text{T}}_p$ and $A^{\text{R}}_q$ is expressed as 
\begin{equation}
\begin{split}
&P'_{p,q,f_i,m_n}(t) =\\& \text{exp}\left(-\tau_{p,q,m_n}(t)\frac{r_\tau-1}{r_\tau DS} \right) 10^{\frac{-Z_n}{10}}\cdot \xi_n(p,q)\cdot \frac{f_i}{f_c}^{\gamma_{m_n}}
\end{split}
\end{equation}
where $Z_n$ is the per cluster shadowing term in dB, $DS$ is the RMS delay spread, $r_\tau$ denotes the delay distribution proportionality factor and determined as the ratio of the standard deviation of the delays to the RMS delay spread \cite{3GPP38901}, $\gamma_{m_n}$ is a environment-dependent random variable and $f_c$ is the center frequency of the communication system. The smooth power variations over the transmit and receive arrays can be simulated by a two-dimension (2D) spatial lognormal process $\xi_n(p,q)$ and can be calculated as 
\begin{equation}
\xi_n(p,q) = 10^{[\mu_n(p,q)+\sigma_n\cdot s_n(p,q)]}
\end{equation}
where $\mu_n(p,q)$ is the local mean and $s_n(p,q)$ is a 2D Gaussian process, which account for the
path loss and shadowing along the large arrays, respectively. The 2D Gaussian process $s_n(p,q)$ is given by 
\begin{equation}
s_n(p,q) = \sum_{k=1}^{K}c_k\cos[(f_{q,k}\delta_q+f_{p,k}\delta_p)+\theta_k]
\end{equation}
where $K$ is the number of sinusoids, $c_k$ are the amplitudes, $f_{q,k}$ and $f_{p,k}$ are the spatial frequencies on both ends. Phases $\theta_k$ are uniformly distributed in $(0,2\pi]$. The final ray powers are obtained by normalizing $P'_{p,q,f_i,m_n}(t)$
\begin{equation}
P_{p,q,f_i,m_n}(t) = \frac{P'_{p,q,f_i,m_n}(t)}{\sum_{n=1}^{N_{p,q}(t)}\sum_{m=1}^{M_{n,f_i}}P'_{p,q,f_i,m_n}(t)}.
\end{equation}

In the scenarios with ultra-massive MIMO, if the channel evolution is taken element by element, the amount of computation will be too large. 
In order reduce the computational costs without the loss of accuracy, the large array can be divided into sub-arrays. In each sub-array, the differences of  total distance for all the elements can be neglected. As a result, the unified array time evolution can be updated only in adjacent sub-arrays. {To determine the size of the sub-array, we calculate the Rayleigh distance \cite{Reyleighdistance} of the sub-array and make sure that the Rayleigh distance is larger than the minimum path distance.} The Rayleigh distance of the sub-array is calculated as $R = 2D_\text{s}^2/\lambda$ where $D_\text{s}$ is the maximum aperture of the sub-array. For example, for a typical indoor THz communication system working at 300 GHz with $1024\times1024$ transmission array, the minimum transmission distance of paths is assumed as 4 m, then the maximum aperture of the sub-array is about 30 mm which supports maximum $60\times 60$ sub-array.

\section{Statistical Properties of the Channel Model}
\subsection{STF-Variant Transfer Function}
The time-variant transfer function $H_{p,q,f_i}(f,t)$ can be calculated as the Fourier transform of the CIR with respect to delay 
\begin{equation}
\begin{split}
&H_{p,q,f_i}(f,t) = \int_{-\infty}^{\infty}h_{p,q,f_i}(f,t)e^{-j2\pi f\tau}d\tau
\\
=&\sqrt{\frac{K^\text{R}}{K^\text{R}+1}}h^\text{L}_{p,q,f_i}(t)e^{-j2\pi f\tau_{p,q}^\text{L}(t)}
\\&+\sqrt{\frac{1}{K^\text{R}+1}}\sum_{n=1}^{N_{p,q}(t)}\sum_{m=1}^{M_{n,f_i}}h^\text{N}_{p,q,f_i,m_n}(t)e^{-j2\pi f\tau_{p,q,m_n}(t)}.
\end{split}
\end{equation}

\subsection{STF Correlation Function}
The STF correlation function (STFCF) measures the correlation properties between two adjacent  channel coefficients
$\rho_{p,q,p',q'}(t,f,\Delta t,\Delta f)$ can be calculated as
\begin{equation}
\label{STFCF}
\begin{split}
&\rho_{p,q,p',q'}(t,f,\Delta t,\Delta f) = 
\\
& E\left[H_{p,q,f_i}(f,t)H^*_{p',q',f'_i}(f+\Delta f, t+\Delta t) \right].
\end{split}
\end{equation}

The (\ref{STFCF}) can be calculated  as the superposition of the  the LOS  and  the NLOS components,~i.e.,
\begin{equation}
\begin{split}
&\rho_{p,q,p',q'}(t,f,\Delta t,\Delta f) =\\&\rho^{LOS}_{p,q,p',p'}(t,f,\Delta t,\Delta f)+\rho^{NLOS}_{p,q,p',p'}(t,f,\Delta t,\Delta f).
\end{split}
\end{equation}
For the Los component, the correlation is expressed as
\begin{equation}
\begin{split}
&\rho^{LOS}_{p,q,p',q'}(t,f,\Delta t,\Delta f) = \\&
\frac{K^\text{R}}{K^\text{R}+1}{h^\text{L}_{p,q,f_i}(t)h^\text{L}_{p',q',f'_i}}^*(t+\Delta t)e^{j2\pi\sigma_1}
\end{split}
\end{equation}
where $\sigma_1=f[\tau_{p,q}^\text{L}(t)-\tau_{p',q'}^\text{L}(t+\Delta t)]+\Delta f\tau_{p',q'}^\text{L}(t+\Delta t)$.
For the NLOS component,  the correlation function  is calculated similarly as
\begin{equation}
\begin{aligned}
&\rho^{NLOS}_{p,q,p',p'}(t,f,\Delta t,\Delta f) =\frac{P_{\text{remain}}(\Delta t,\tilde{\delta}_{p}, \tilde{\delta}_{q} )}{K^\text{R}+1}
\\
&E\left[\sum_{n=1}^{N_{p,q}(t)}\sum_{n'=1}^{N_{p',q'}(t)}\sum_{m=1}^{M_{n,f_i}}\sum_{m'=1}^{M_{n,f'_i}}P_{\text{remain}}(\Delta f)\right.
\\
&\phantom{=\;\;}\left.h^\text{N}_{p,q,f_i,n_m}(t){h^\text{N}_{p',q',f'_i,{n'}_{m'}}}^*(t+\Delta t)e^{j2\pi\sigma_2}\right]
\end{aligned}
\end{equation}

where $\sigma_2=f[\tau_{p,q,n_m}(t)-\tau_{p',q',n_m}(t+\Delta t)]+\Delta f\tau_{p',q',n_m}(t+\Delta t)$.
It is hard to present the STFCF directly because of  the high dimension of the STFCF. However, by setting $\Delta f = 0$, $p = p'$, and $q = q'$, the STFCF is reduced to the time-frequency-variant ACF
\begin{equation}
\begin{split}
\rho_{p,q}^{\text{ACF}}(t, f, \Delta t) &= \rho_{p,q}(t,f,\Delta t,0)\\&=E\left[H_{p,q,f_i}(f,t)H^*_{p,q,f_i}(f, t+\Delta t) \right].
\end{split}
\end{equation}
By setting $\Delta f = 0$ and $\Delta t = 0$, STFCF is reduced to the space cross-correlation function (SCCF)
\begin{equation}
\begin{split}
\rho_{p,q,p',q'}^{\text{SCCF}}(t, f) &= \rho_{p,q,p',q'}(t,f,0,0)\\&=E\left[H^*_{p,q,f_i}(f,t)H_{p',q',f_i}(f, t) \right].
\end{split}
\end{equation}
Similarly, by setting $\Delta t = 0$, $p = p'$, and $q = q'$, the STFCF is reduced to the time-frequency-variant frequency correlation function (FCF)
\begin{equation}
\begin{split}
\rho_{p,q}^{\text{FCF}}(t, f, \Delta f) &= \rho_{p,q}(t,f,0,\Delta f)
\\&
=E\left[H^*_{p,q,f_i}(f,t)H_{p,q,f_i}(f+\Delta f, t) \right].
\end{split}
\end{equation}
\subsection{Delay PSD and Delay Spread}
The time-frequency variant  delay PSD for the  channel between the $p$th Tx antenna and the $q$th Rx antenna can be expressed as
\begin{equation}
\Upsilon_{p,q}(t,f_{i},\tau) = \sum_{n=1}^{N_{p,q}(t)}\sum_{m=1}^{M_{n,f_i}}\left| h_{p,q,f_i,m_n}(t)\right| ^2\delta(\tau-\tau_{p,q,m_n}(t)).
\end{equation}
The time-variant delay PSD is influenced by the time-dependent mean powers of rays with clusters and delays of rays parameters. 

The delay spread is an important parameter in determining whether or not an adaptive equalizer is required at the receiver. If the delay spread exceeds 10 percent to 20 percent of the symbol duration, then an adaptive equalizer is required. The delay spread of the channel model can be written as 
\begin{equation}
  \begin{split}
   &\sigma_{\tau}(t,f_i) = \\&\sqrt{\frac{\sum\limits_{n=1}^{N_{p,q}(t)}\sum\limits_{m=1}^{M_{n,f_i}}(\tau_{p,q,m_n}(t)-\mu_\tau)^2\Upsilon_{p,q}(t,f_{i},\tau_{p,q,m_n}(t))}{\sum\limits_{n=1}^{N_{p,q}(t)}\sum\limits_{m=1}^{M_{n,f_i}}\Upsilon_{p,q}(t,f_{i},\tau_{p,q,m_n}(t))}}.
  \end{split}
\end{equation}
where $\mu_\tau$ is the average delay of the channel and is calculated as 
\begin{equation}
\mu_\tau = \frac{\sum\limits_{n=1}^{N_{p,q}(t)}\sum\limits_{m=1}^{M_{n,f_i}}(\tau_{p,q,m_n}(t)-\mu_\tau)\Upsilon_{p,q}(t,f_{i},\tau_{p,q,m_n}(t))}{\sum\limits_{n=1}^{N_{p,q}(t)}\sum\limits_{m=1}^{M_{n,f_i}}\Upsilon_{p,q}(t,f_{i},\tau_{p,q,m_n}(t))}
\end{equation}

\subsection{Local Doppler PSD}
The local Doppler PSD measures the average power  as a function of the Doppler frequency. It  can be calculated by  the Fourier transform of the temporal ACF with respect to $\Delta t$ and can be written as
\begin{equation}
S_{p,q}(f,t) = \int_{-\infty}^{\infty}\rho_{p,q}^{\text{ACF}}(t, f, \Delta t)e^{-j2\pi f\Delta t}d\Delta t.
\end{equation}

\subsection{Angular PSD and Angular Spread}
The angular PSD of the channel model measures the power distribution in the angular domain. It can be calculated from the CIR of the channel model and written as 
\begin{equation}
\begin{split}
&\Lambda^\text{T}(t,\Omega_E,\Omega_A) \\=& \sqrt{\frac{K^\text{R}}{K^\text{R}+1}}\left| h^\text{L}_{p,q,f_i}(t)\right|^2 \delta(\Omega_E-\phi^{\text{T}}_{E,L}(t))\delta(\Omega_A-\phi^{\text{T}}_{A,L}(t))
\\
&+\sqrt{\frac{1}{K^\text{R}+1}}\sum_{n=1}^{N_{p,q}(t)}\sum_{m=1}^{M_{n,f_i}}\left| h^\text{N}_{p,q,f_i,m_n}(t)\right|\delta(\Omega_E-\phi^{\text{T}}_{E,m_n}(t))\\
&\delta(\Omega_A-\phi^{\text{T}}_{A,m_n}(t))
\end{split}
\end{equation}	
where $\Omega_E$ and $\Omega_A$ are the EAoD and AAoD at the Tx side. The angular PSD at the Rx side can be calculated in the similar method.

The cluster level angle spread reflects degree of the diffusely scattering  and can also be obtained from the CIR. The cluster level angle spread $\psi^\text{T}_{A,n}$ is expressed as 
\begin{equation}
\psi^\text{T}_{A,n,f_i}(t) = \sqrt{\frac{\sum_{m=1}^{M_{n,f_i}}(\Delta \phi^{\text{T}}_{A,m_n})^2\Lambda^\text{T}(t,\phi^{\text{T}}_{A,m_n})}{\sum_{m=1}^{M_{n,f_i}}\Lambda^\text{T}(t,\phi^{\text{T}}_{A,m_n})}}.
\end{equation}

\subsection{Stationary Regions in STF Domains}

For non-wide sense stationary channel, stationary interval is utilized in \cite{JCOM12_Stationarityintervals} to measure the  period within which the channel amplitude response can be seen as stationary.
In this model, non-stationarities in space, time, frequency domains are considered. Therefore,  an improved definition for the stationary regions is extended to space and frequency domains.
The stationary regions in space, time, and frequency domains are denoted as stationary distance, stationary interval, and stationary bandwidth and can be calculated  as maximum distance/interval/frequency bandwidth within which the auto-correlation function of time-frequency dependent delay PSD exceeds the threshold. 

The stationarity regions in three domains can be  written as 
\begin{equation}
\begin{split}
&I(\Delta p, \Delta q, \Delta t, \Delta f)=
\\&\text{inf}\left\lbrace \Delta p, \Delta q, \Delta t, \Delta f\mid_{R_{\Upsilon_{p,q,p',q'}(t,f,\Delta p, \Delta q,\Delta t, \Delta f)}<c_{th}}\right\rbrace
\end{split}
\end{equation}
where $\text{inf} \left\lbrace \cdot \right\rbrace $
calculates the infimum of a function . It should be noticed that the  threshold $c_{th}$  can be adjusted for different requirements.  The  normalized ACF of the delay PSD $R_{\Upsilon_{p,q,p',q'}(t,f,\Delta p, \Delta q,\Delta t, \Delta f)}$ is defined by
\begin{equation}\label{stationary_bandwidth}
\begin{split}
&R_{\Upsilon_{p,q,p',q'}(t,f,\Delta p, \Delta q,\Delta t, \Delta f)}=\\&\frac{\int \Upsilon_{p,q}(t,f_{i},\tau)\Upsilon_{p',q'}(t_i+\Delta t,f_{i}+\Delta f,\tau)d\tau}{{max}\left\lbrace \int \Upsilon_{p,q}^2(t,f_{i},\tau) d\tau, \int \Upsilon_{p',q'}^2(t_i+\Delta t,f_{i}+\Delta f,\tau) d\tau\right\rbrace}
\end{split}
\end{equation}
where $\Upsilon_{p,q}(t,\tau,f_{i})$ is the delay PSD of the model and $p' = p+\Delta p$, $q' = q+\Delta q$. 

For the time domain, by setting the $\Delta p $ = 0, $\Delta q $ = 0, $\Delta f $ = 0 in (\ref{stationary_bandwidth}), the stationary interval is written as 
\begin{equation}
I(\Delta t)=\text{inf}\left\lbrace \Delta t \mid_{R_{\Upsilon_{p,q}(t,f,\Delta t)}<c_{th}}\right\rbrace.
\end{equation}
For the space domain, the stationary distance can be obtained by setting  $\Delta t $ = 0, $\Delta f $ = 0 in (\ref{stationary_bandwidth}) as
\begin{equation}
I(\Delta p,\Delta q)=\text{inf}\left\lbrace \Delta p,\Delta q \mid_{R_{\Upsilon_{p,q,p',q'}(t,f)}<c_{th}}\right\rbrace.
\end{equation}
For the frequency  domain, the stationary  bandwidth can be calculated  by setting the $\Delta p $ = 0, $\Delta q $ = 0, $\Delta t $ = 0  as 
\begin{equation}
I(\Delta f)=\text{inf}\left\lbrace \Delta f \mid_{R_{\Upsilon_{p,q}(t,f,\Delta f)}<c_{th}}\right\rbrace.
\end{equation}


\section{Results and Analysis}
In this section,  the statistical properties of the THz channel models are simulated and discussed. The frequency of the system  is chosen from 300~GHz to 350~GHz. The Tx array is set as 128$\times$128 and The Rx array is set as 16$\times$16. Both $\delta_\text{T}$ and $\delta_\text{R}$ equal to half of wavelength at 325 GHz. {To validate the proposed THz channel model, basic simulation parameters for a typical THz indoor scenario are set as follows.} The elevation and azimuth angles of Tx array $\beta_{V,E}^\text{T}$, $\beta_{V,A}^\text{T}$, $\beta_{H,E}^\text{T}$, and $\beta_{H,A}^\text{T}$ are set as $\frac{-\pi}{10}$, $\frac{\pi}{10}$, $\frac{-3\pi}{5}$, and $\frac{3\pi}{4}$, respectively. The velocity of Rx is $\text{v}^\text{R}$ = 0.6 m/s  with  $\alpha^\text{R}_E$ = 0 and $\alpha^\text{R}_A$~=~$\frac{\pi}{3}$. The Tx is assumed as fixed. The bandwidth of sub-band is 0.1 GHz.

\subsection{STF Correlation Function}
\begin{figure}[t]
	\centerline{\includegraphics[width=0.5\textwidth]{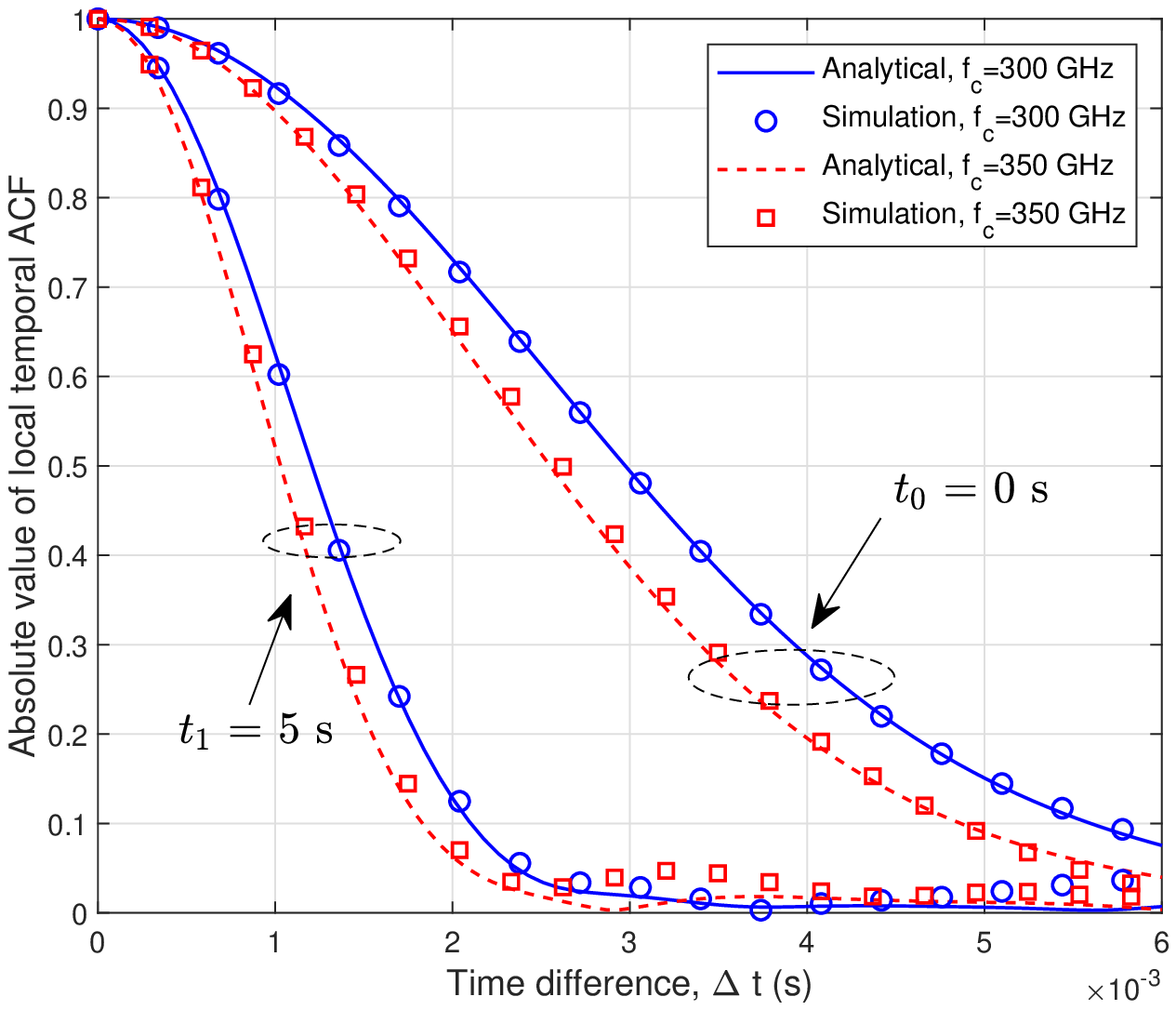}}
	\caption{{The comparison of time-variant ACFs for cluster $C_1^\text{T}$ at $t_0$ = 0 s and $t_1$ = 5 s at 300 GHz and 350 GHz ($D = 3$ m, $d_{p,q,n} = 5 $ m,  $p$ = 1, $q$ = 1, $\text{v}^\text{R}$ = 0.6 m/s, $\text{v}^\text{T}$ = 0 m/s, $\sigma_{A,n}$ = $2.8^\circ$,  $\sigma_{E,n}$ = $1.4^\circ$).}}
	\label{fig_ACF}
\end{figure}

The simulations for correlation functions are assumed to imitate indoor scenarios. {The analytic results are obtained by ideal cases where the paths in each cluster are trend to infinity. However, it is unprocurable in the real hardware. The simulated results are calculated after the channel generation. It can be utilized in the real communication hardware.}  The generation rate $\lambda_{G}$ is assumed as 0.8, and the recombination rate $\lambda_{R} = 0.04$,  the mean number of clusters is 20. The time ACF of the model can be calculated by setting $\Delta p$, $\Delta q$, and $\Delta f$ as 0.
The comparison of channel model at  $t_0$ = 0 s and $t_1$ = 5 s of {${C}_1^{\text{T}}$}  is shown in Fig.~\ref{fig_ACF}.  We can observe from the figure that the simulation results fit well with the analytical results. The difference of ACFs at different instants and frequencies can be obviously observed verifying that  the non-stationarities caused by mobility and large bandwidth are nonnegligible.
\begin{figure}[t]
	\centerline{\includegraphics[width=0.5\textwidth]{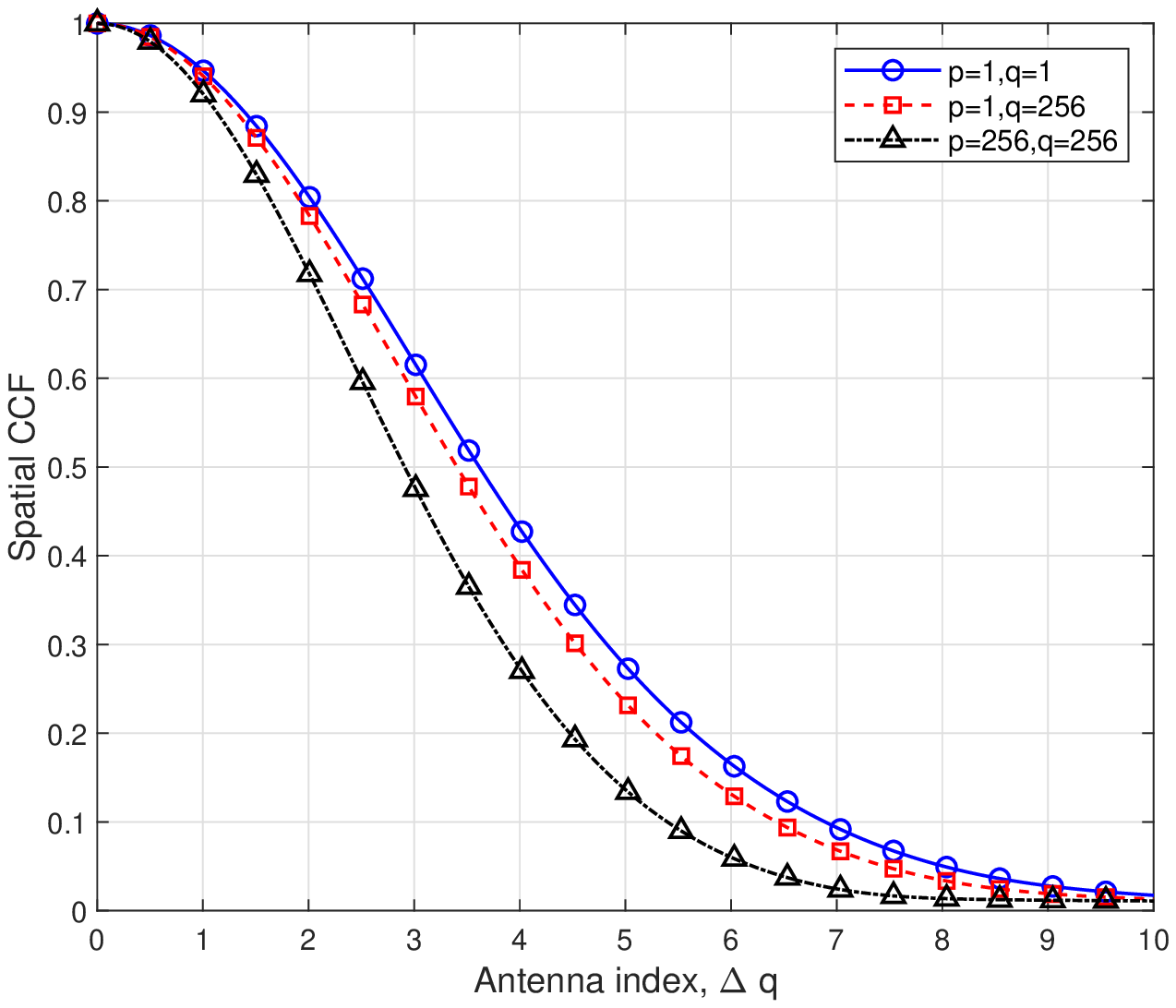}}
	\caption{The comparison of spatial CCFs at different antenna at different antenna element ($D = 3$ m, $f_i = 300$ GHz, $\text{v}^\text{R}$ = 0.6 m/s, $\text{v}^\text{T}$ = 0 m/s, $\sigma_{A,n}$ = $2.8^\circ$,  $\sigma_{E,n}$ = $1.4^\circ$).}
	\label{fig_SCCF}
\end{figure}

\begin{figure}[t]
	\centerline{\includegraphics[width=0.5\textwidth]{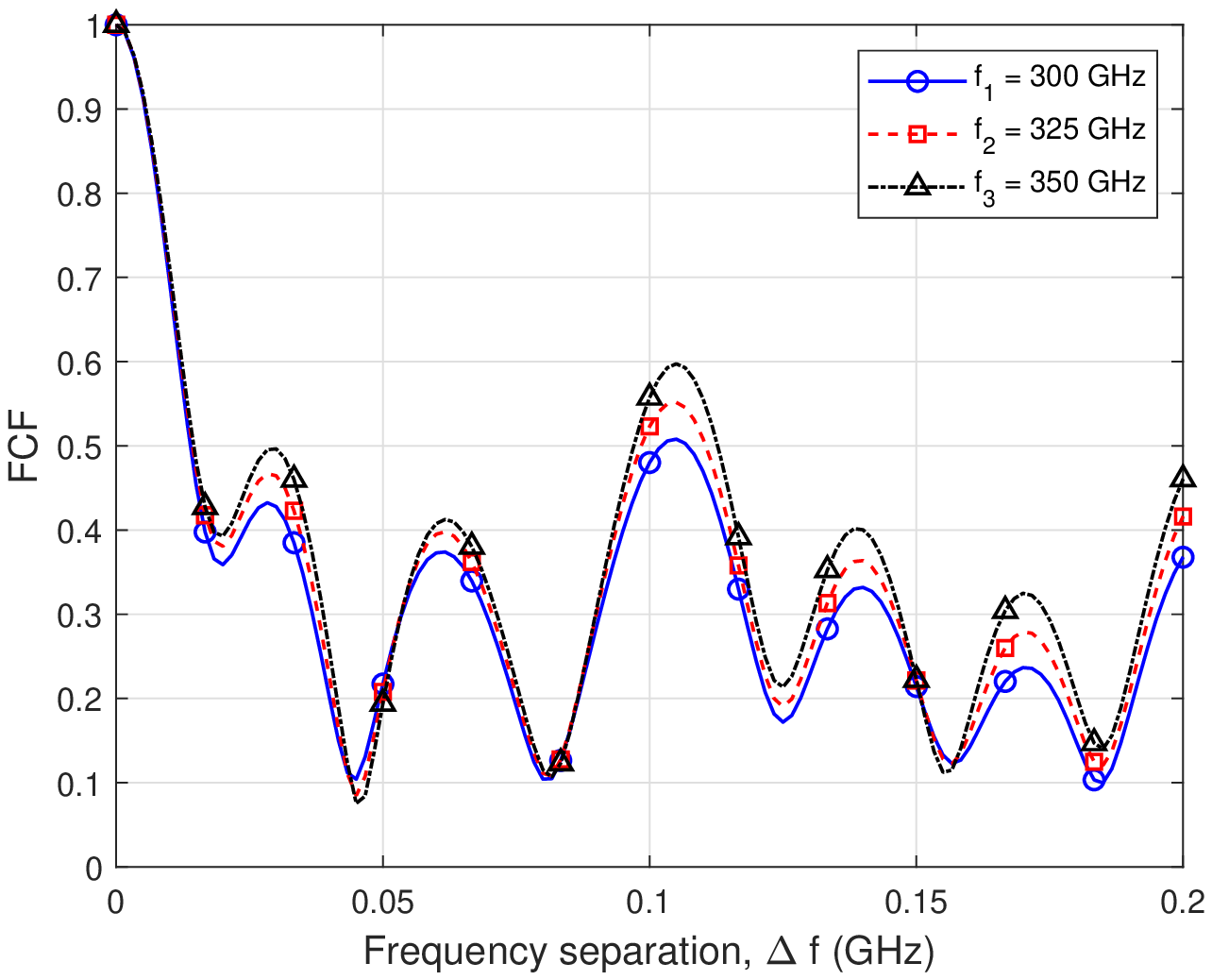}}
	\caption{The comparison of FCFs at different frequencies ($D = 3$ m,  $p$ = 1, $q$ = 1,  $\text{v}^\text{R}$ = 0.6 m/s, $\text{v}^\text{T}$ = 0 m/s, $\sigma_{A,n}$ = $2.8^\circ$,  $\sigma_{E,n}$ = $1.4^\circ$).}
	\label{fig_fcf}
\end{figure}

The spatial CCFs for {${C}_1^{\text{T}}$} at different elements are demonstrated in Fig. \ref{fig_SCCF}. { We can observe that the differences for  spatial CCFs at different elements are nonnegligible which verifies the spatial non-stationarity caused by ultra-massive MIMO. }

The absolute values of FCFs for NLOS paths at different  $f_1$ = 300 GHz, $f_2$ = 325 GHz, and $f_3 = 350$ GHz for THz massive MIMO channel model are shown in Fig.~\ref{fig_fcf}. We can notice that differences among FCFs at different frequencies are small but still observable which means that  the non-stationarity in the frequency domain needs to be considered.

\subsection{Stationary Bandwidth}
\begin{figure}[t]
	\centerline{\includegraphics[width=0.5\textwidth]{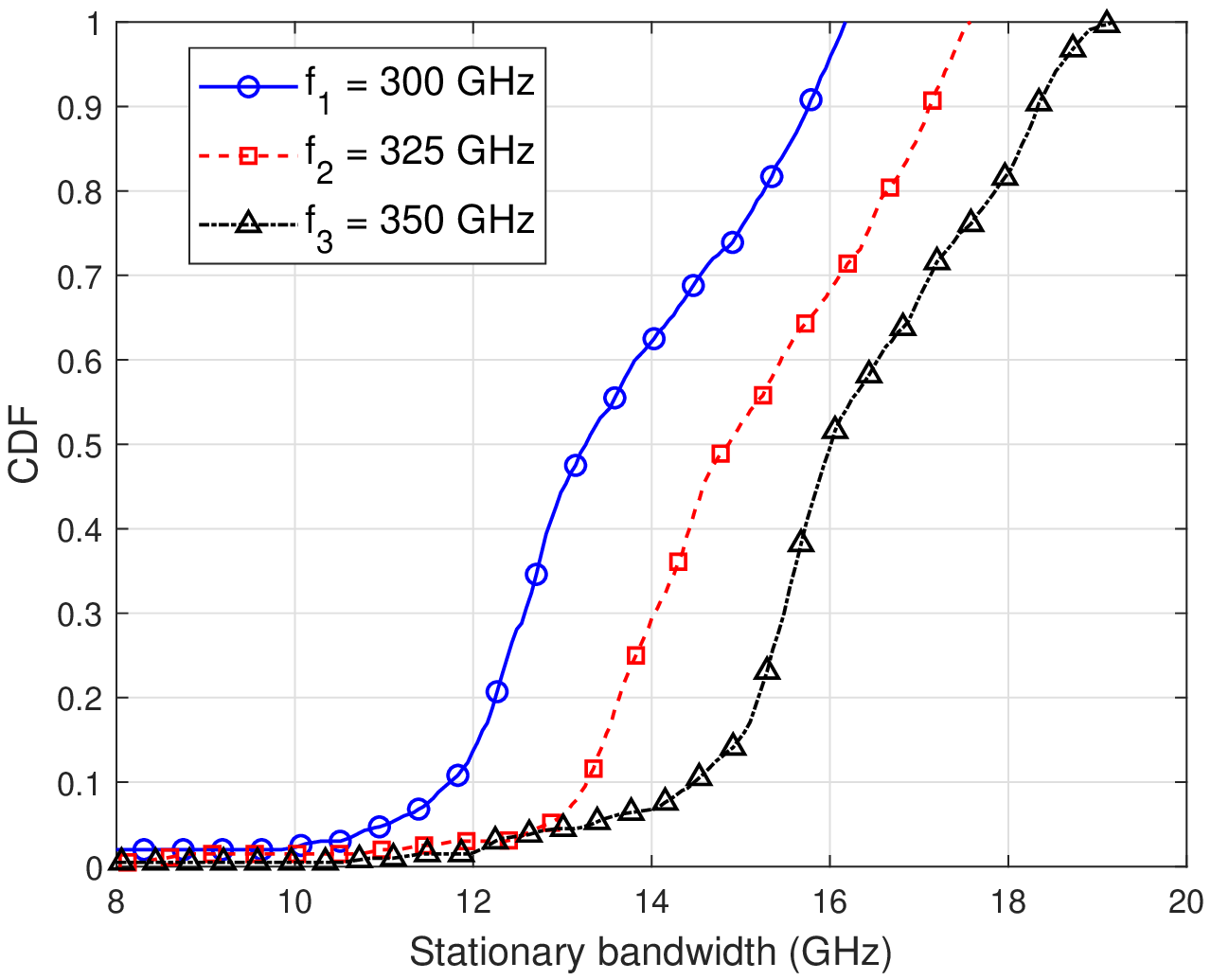}}
	\caption{The comparison of stationary bandwidth at different frequencies ($D = 3$ m, $\bar{d}^\text{N} = 1.5 $ m,  $p$ = 1, $q$ = 1, $\text{v}^\text{R}$ = 1 m/s, $\text{v}^\text{T}$ = 0 m/s, $\sigma^\text{R}_{A,n}$ = $1.7^\circ$,  $\sigma^\text{R}_{E,n}$ = $1.2^\circ$, $\sigma^\text{T}_{A,n}$ = $2.8^\circ$,  $\sigma^\text{T}_{E,n}$ = $1.4^\circ$).}
	\label{stationaryintervalg}
\end{figure}

The stationary bandwidth is simulated with parameters of a typical indoor scenario according to\cite{RN194,RN504}. The initial distance between the first elements of Tx and Rx is 3 m, $\bar{d}^\text{N}$ is set as 1.5 m. 
The initial intra-cluster parameters $\sigma_{E,n}^\text{R}$, $\sigma_{A,n}^\text{R}$,  $\sigma_{E,n}^\text{T}$, and $\sigma_{A,n}^\text{T}$ are set as $1.2^\circ$, $1.7^\circ$, $1.4^\circ$, and $2.8^\circ$, respectively. 
The  cumulative distribution functions (CDFs) of the stationary bandwidth of the simulation channel model at different frequencies are shown in Fig.~\ref{stationaryintervalg}. 
The threshold is chosen as 0.9. 
The median of the stationary bandwidth at 300 GHz is approximately 12.5~GHz. Higher frequency band has larger frequency stationary bandwidth. If the bandwidth in THz communication is larger than the stationary bandwidth, non-stationarity in frequency domain is nonnegligible.


\subsection{Cluster Level Angle Spread}

\begin{figure}[t]
	\centerline{\includegraphics[width=0.5\textwidth]{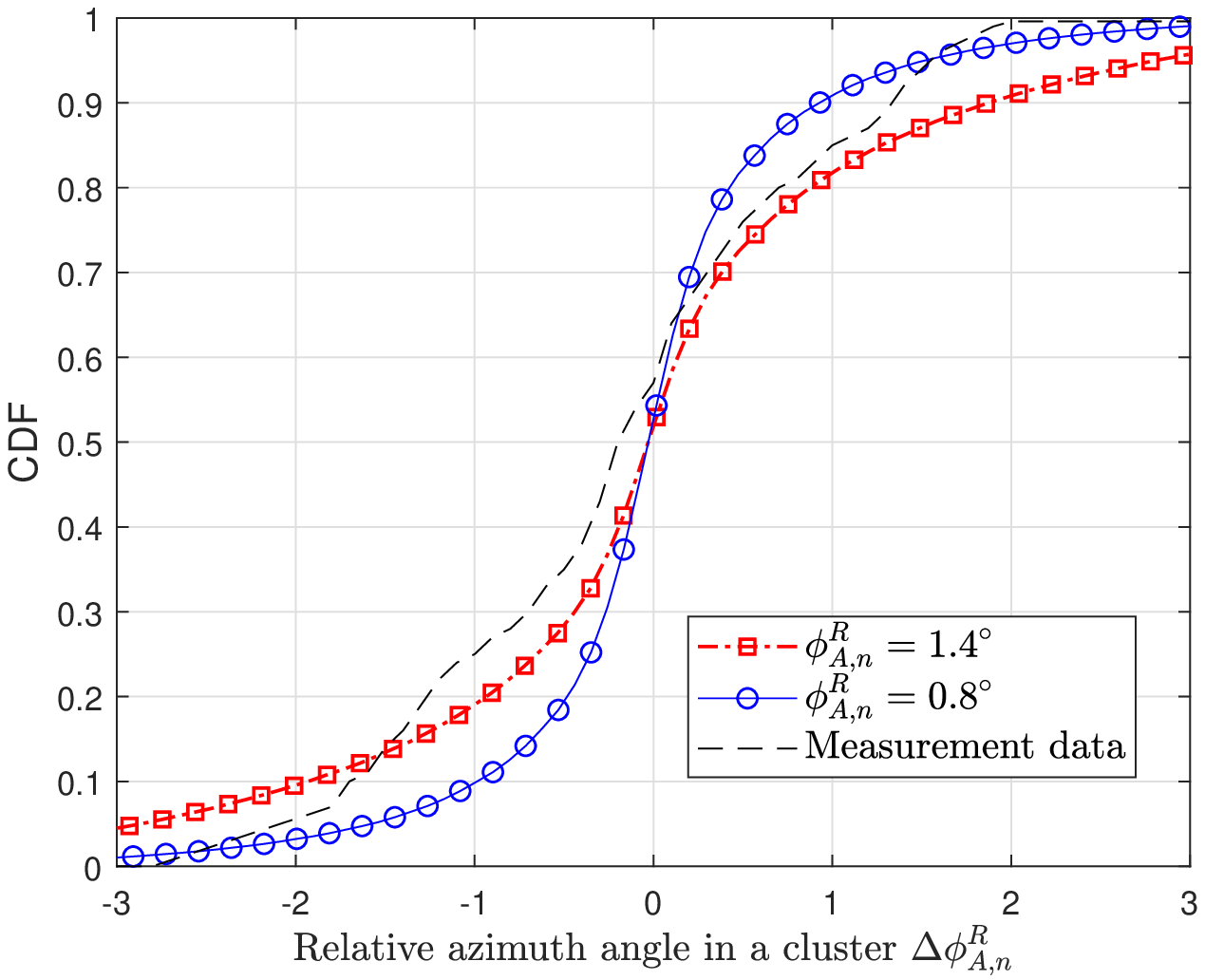}}
	\caption{The CDFs of relative angle in a cluster with different $\sigma_{E,n}^\text{R}$ and the measurement in \cite{RN197} ($D = 2.7$ m, $p$ = 1, $q$ = 1, $t_0$ = 0 s, $f_0$ = 300 GHz, $\text{v}^\text{R}$ = 0 m/s).}
	\label{cdfclusterangle}
\end{figure}
\begin{figure}[t]
	\centerline{\includegraphics[width=0.5\textwidth]{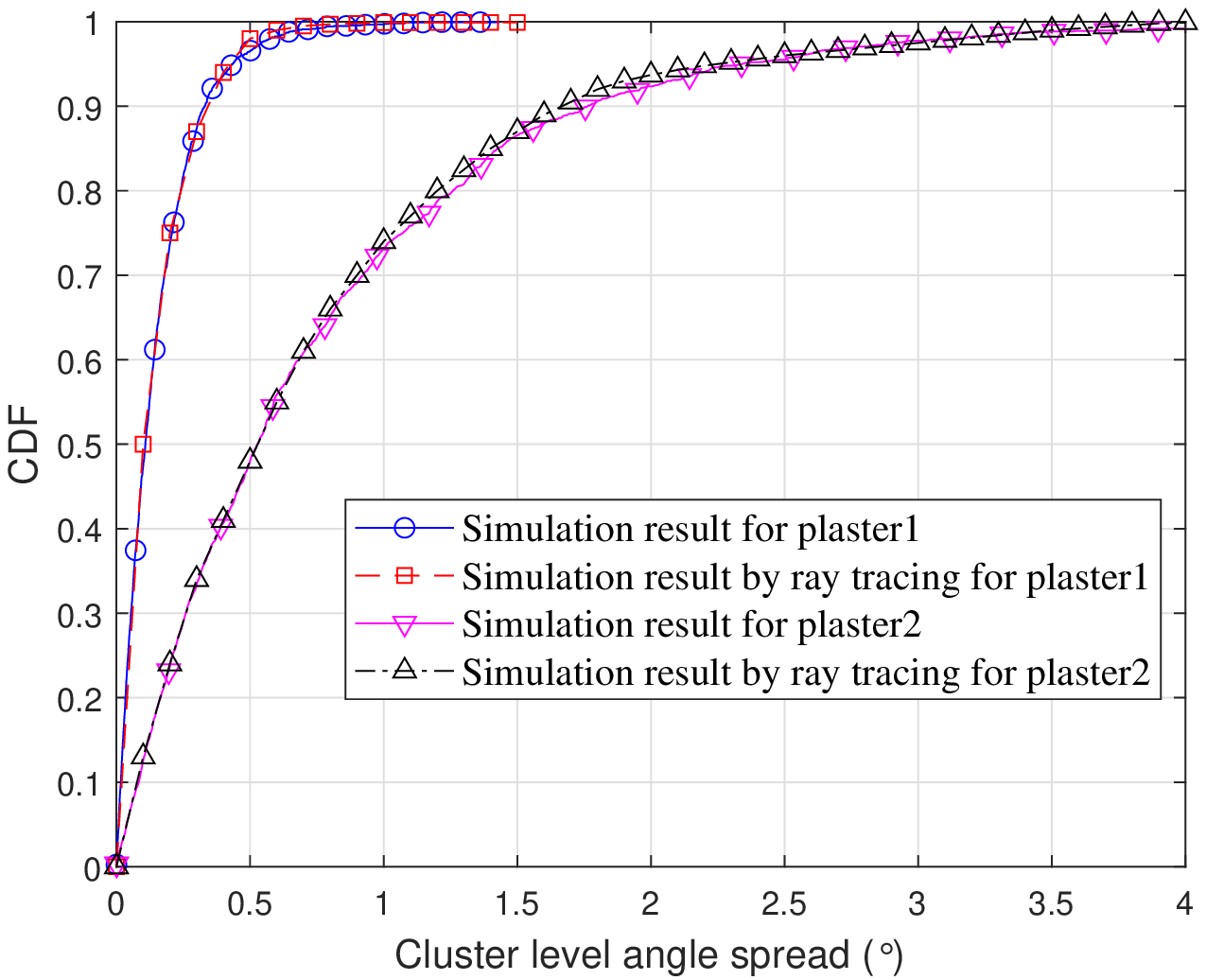}}
	\caption{{The CDFs of cluster  angle spread with different materials  ($D = 2.7$ m, $p$ = 1, $q$ = 1, $t_0$ = 0 s, $f_0$ = 300 GHz, $\sigma_{A,n}^{\text{R}}$ = $0.15^\circ$ for  plaster1, $\sigma_{A,n}^{\text{R}}$ = $0.75^\circ$ for  plaster2).}}
	\label{clusterangle}
\end{figure}
Cluster level angle spread reflects the degree of diffusely scattering in a cluster. 
In this simulation, the parameter $\sigma_{E,n}^\text{R}$ 
is optimized according to minimum mean square error (MMSE) criterion to fit the data from the channel measurements \cite{RN197} or ray tracing \cite{RN195}. In the optimization process,  an iterative process is conducted to determine the minimum mean square error and its corresponding parameters.  
The CDFs of relative azimuth angles of arrival with different $\sigma_{E,n}^\text{R}$ are simulated and compared with the measurement data\cite{RN197} in Fig. \ref{cdfclusterangle}. {The measurements in \cite{RN197} were conducted in a small static office scenario with some furniture. The measured frequency was from 275 GHz to 325 GHz. The distance between Tx and Rx is 2.8~m.} We can observe the good agreement when $\sigma_{E,n}^\text{R}$=$1.4^\circ$. Then we simulate the scattering in two materials with different roughness. The cluster level angle spread is greatly affected by the roughness of the material. The height standard deviation of plaster1 is $\sigma_{\text{plaster1}}=0.5$ mm and $\sigma_{\text{plaster2}}=1.5$~mm in \cite{RN195}.  Fig.~\ref{clusterangle} demonstrates the comparison of cluster level angle spread with different materials in this  model and simulation by ray tracing in \cite{RN195}. The results show that the proposed  model can approximate well with the ray tracing for different materials.

\subsection{RMS Delay Spread}

\begin{figure}[t]
	\centerline{\includegraphics[width=0.5\textwidth]{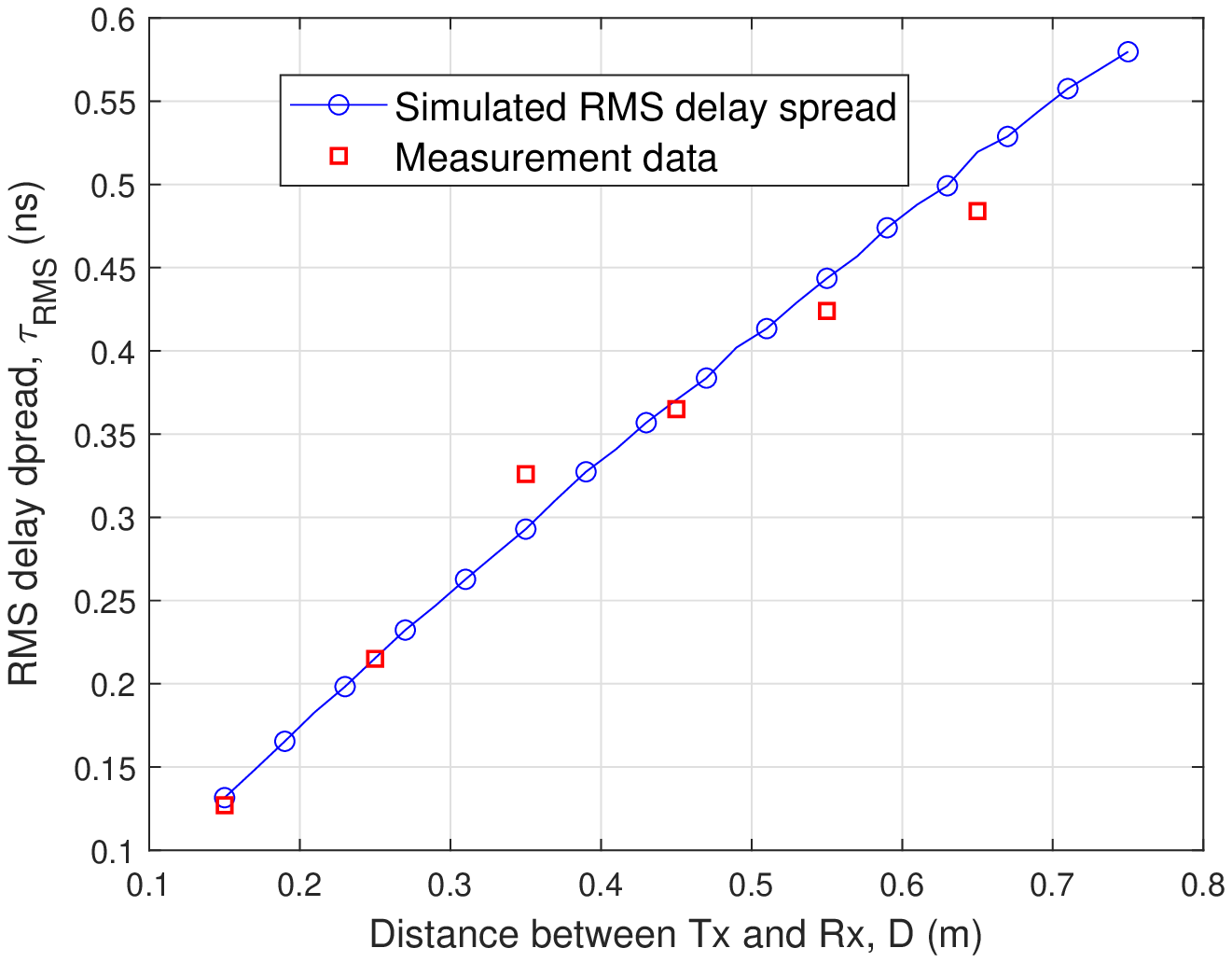}}
	\caption{The comparison of simulated RMS delay spread and measurement in \cite{RN448} ($p$ = 1, $q$ = 1, $t_0$ = 0 s, $f_0$ = 300~GHz, $\text{v}^\text{R}$ = 0~m/s, $\sigma_{E,n}^\text{R}$ = 0.1,  $\sigma_{E,n}^\text{R}$ = 0.07, $\bar{d}^\text{N}$ = 0.05 m for $D$ = 0.15 m).}
	\label{rms}
\end{figure}
The RMS delay spread is simulated and fitted with the measurement data\cite{RN448} by minimum mean square error criterion in Fig. \ref{rms}. {In\cite{RN448}, channel measurement on the desktop where the transmission distance is less than 1 m at 300 GHz was conducted.} In the simulation, the transmission distance is from 0.1 m to 0.75 m and a single antenna for both the Tx and Rx is applied. The parameters $\lambda_{G} $ = 0.2 and $\lambda_{R} $ = 0.04. $\bar{d}^\text{N}$ are linearly increase with the LOS distance and the parameters at $D$ = 0.15 m are estimated. We can observe good agreement between the simulation results and the measurement data. It means that the proposed model is suitable for this scenario. 

\section{Conclusions}

A novel 3D general THz GBSM for 6G wireless communication systems has been  proposed in this paper. The non-stationarities and cluster evolution  in space, time, and frequency domains based on birth-death processes have been demonstrated. The novel general THz channel model can be reduced to specific channel models to fit various scenarios by adjusting  model parameters.  Statistical properties of the proposed general THz channel model such as the correlation functions and stationarity regions  have been derived and simulated. The good agreements between simulated statistical properties and corresponding channel measurements have verified the accuracy and generality of the proposed THz model. The proposed general THz channel model can be utilized in 6G THz communication systems design for different scenarios with good accuracy.

\begin{IEEEbiography}[{\includegraphics[width=1in,clip,keepaspectratio]{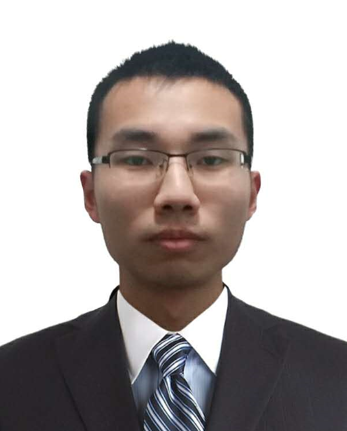}}]{Jun Wang}
	received the B.E. degree in Information Engineering from Southeast University, China, in 2016. He is currently pursuing the Ph.D. degree in the Nation Mobile
	Communications Research Laboratory, Southeast University, China. His research interests is THz wireless channel measurements and modeling.
\end{IEEEbiography}

\begin{IEEEbiography}[{\includegraphics[width=1in,height=1.25in,clip,keepaspectratio]{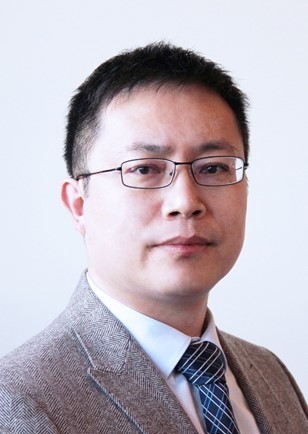}}]{Cheng-Xiang Wang}
	(S'01-M'05-SM'08-F'17) received the BSc and MEng degrees in
	Communication and Information Systems from	Shandong University, China, in 1997 and 2000,	respectively, and the PhD degree in Wireless Communications	from Aalborg University, Denmark, in 2004.
	
	He was a Research Assistant with the Hamburg University of Technology, Hamburg, Germany, from 2000 to 2001, a Visiting Researcher with
	Siemens AG Mobile Phones, Munich, Germany, in 2004, and a Research
	Fellow with the University of Agder, Grimstad, Norway, from 2001 to 2005. He has been with Heriot-Watt University, Edinburgh, U.K., since 2005, where he was promoted to a Professor in 2011. In 2018, he joined the National Mobile Communications Research Laboratory, Southeast University, China, as a Professor. He is also a part-time professor with the Purple Mountain Laboratories, Nanjing, China. He has authored four books, three book chapters, and more than 410 papers in refereed journals and conference proceedings, including 24 Highly Cited Papers. He has also delivered 22 Invited Keynote Speeches/Talks and~7 Tutorials in international conferences. His current research interests include wireless channel measurements and modeling, 6G wireless communication networks, and applying artificial intelligence to wireless networks.
	
	Prof. Wang is a member of the Academia Europaea, a fellow of the IET, an IEEE Communications Society Distinguished Lecturer in 2019 and 2020, and a Highly-Cited Researcher recognized by Clarivate Analytics in 2017-2020. He is currently an Executive	Editorial Committee member for the IEEE TRANSACTIONS ON WIRELESS COMMUNICATIONS. He has served as an Editor for nine international journals, including the IEEE TRANSACTIONS ON WIRELESS
	COMMUNICATIONS from 2007 to 2009, the IEEE TRANSACTIONS
	ON VEHICULAR TECHNOLOGY from 2011 to 2017, and the
	IEEE TRANSACTIONS ON COMMUNICATIONS from 2015 to 2017.
	He was a Guest Editor for the IEEE JOURNAL ON SELECTED AREAS
	IN COMMUNICATIONS, Special Issue on Vehicular Communications and
	Networks (Lead Guest Editor), Special Issue on Spectrum and Energy
	Efficient Design of Wireless Communication Networks, and Special Issue
	on Airborne Communication Networks. He was also a Guest Editor for 	the IEEE TRANSACTIONS ON BIG DATA, Special Issue on Wireless 	Big Data, and is a Guest Editor for the IEEE TRANSACTIONS ON 	COGNITIVE COMMUNICATIONS AND NETWORKING, Special Issue on Intelligent Resource Management for 5G and Beyond. He has served as a TPC Member, a TPC Chair, and a General Chair for more than 80 international conferences. He received 12 Best Paper Awards from IEEE GLOBECOM 2010, IEEE ICCT 2011, ITST 2012, IEEE VTC 2013-Spring, IWCMC 2015, IWCMC 2016, IEEE/CIC ICCC 2016, WPMC 2016, WOCC 2019, IWCMC 2020, and WCSP 2020.
\end{IEEEbiography}

\begin{IEEEbiography}[{\includegraphics[width=1in,height=1.25in,clip,keepaspectratio]{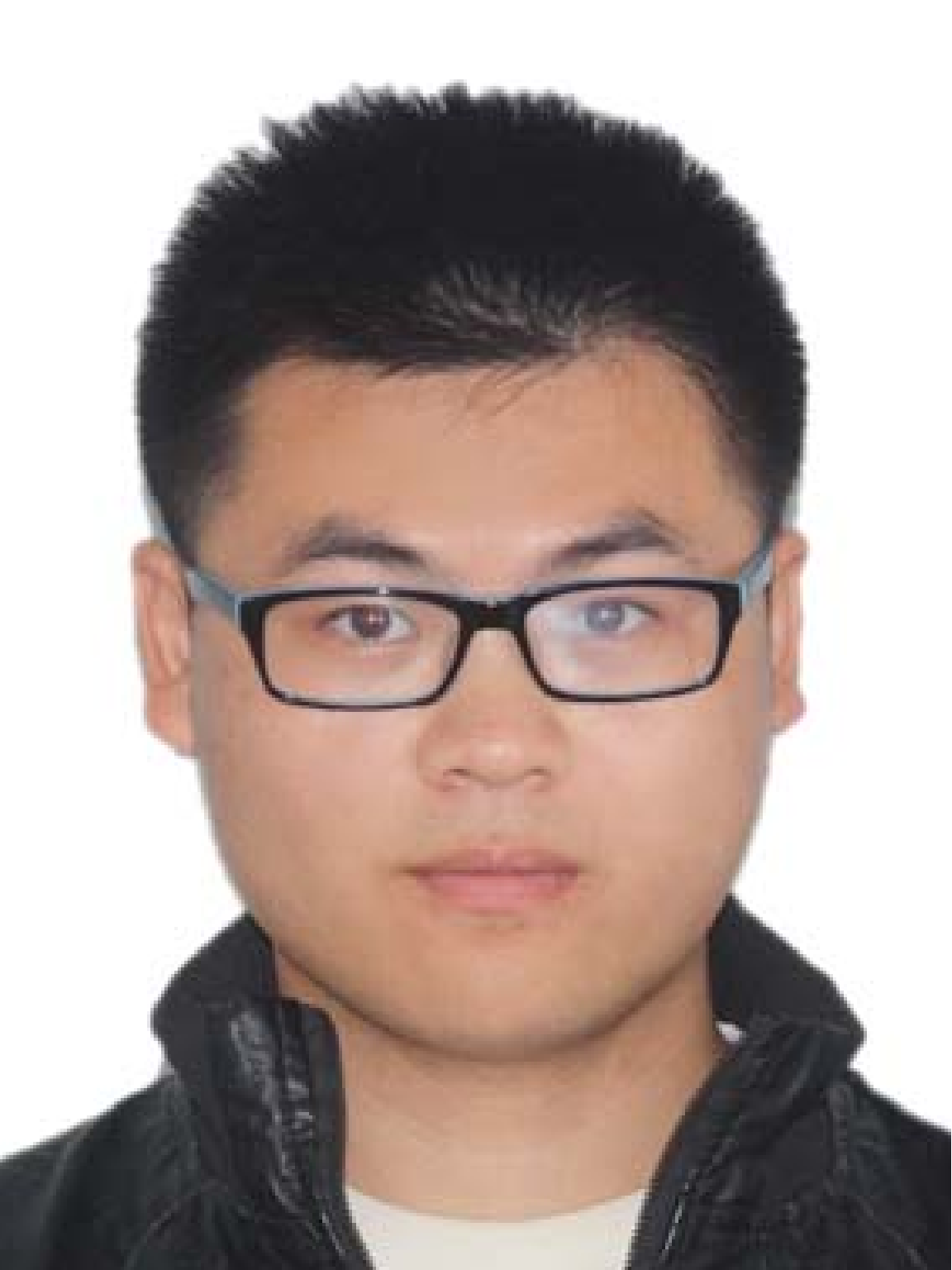}}]{Jie Huang}
	(M'20) received the B.E. degree in information engineering from Xidian University, China, in 2013, and the Ph.D. degree in communication and information systems from Shandong University, China, in 2018. From October 2018 to October 2020, he was a Research Associate with the National Mobile Communications Research Laboratory, Southeast University, China. From January 2019 to February 2020, he was a Research Associate with Durham University, U.K. He is currently an Associate Professor with the National Mobile Communications Research Laboratory, Southeast University, China and a Researcher with the Purple Mountain Laboratories, China. His research interests include millimeter wave, THz, massive MIMO, and intelligent reflecting surface channel measurements and modeling, wireless big data, and 6G wireless communications.
\end{IEEEbiography}

\begin{IEEEbiography}
	[{\includegraphics[width=1.1in,height=1.3in,clip,keepaspectratio]{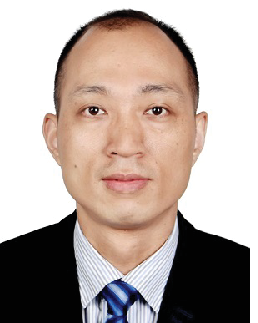}}]
	{Haiming Wang} (Member, IEEE) was born in 1975. He received the B.S., M.S., and Ph.D. degrees in Electrical Engineering from Southeast University, Nanjing, China, in 1999, 2002, and 2009, respectively.
	
	He joined the State Key Laboratory of Millimeter Waves and the School of Information Science and Engineering, Southeast University, Nanjing, China, in 2002, and is currently a Distinguished Professor. He is also a part-time professor with the Purple Mountain Laboratories, Nanjing, China. He has authored and co-authored over 50 technical publications in IEEE TRANSACTIONS ON ANTENNAS AND PROPAGATION and other peer-reviewed academic journals. Prof. Wang has authored and co-authored over more than 70 patents and 52 patents have been granted. He was awarded for contributing to the development of IEEE 802.11aj by the IEEE Standards Association in July 2018.
	
	His current research interests include AI-powered antenna and radiofrequency technologies, AI-powered channel measurement and modeling technologies, and millimeter-wave and THz wireless communications.
\end{IEEEbiography}

\begin{IEEEbiography}[{\includegraphics[width=1in,height=1.25in,clip,keepaspectratio]{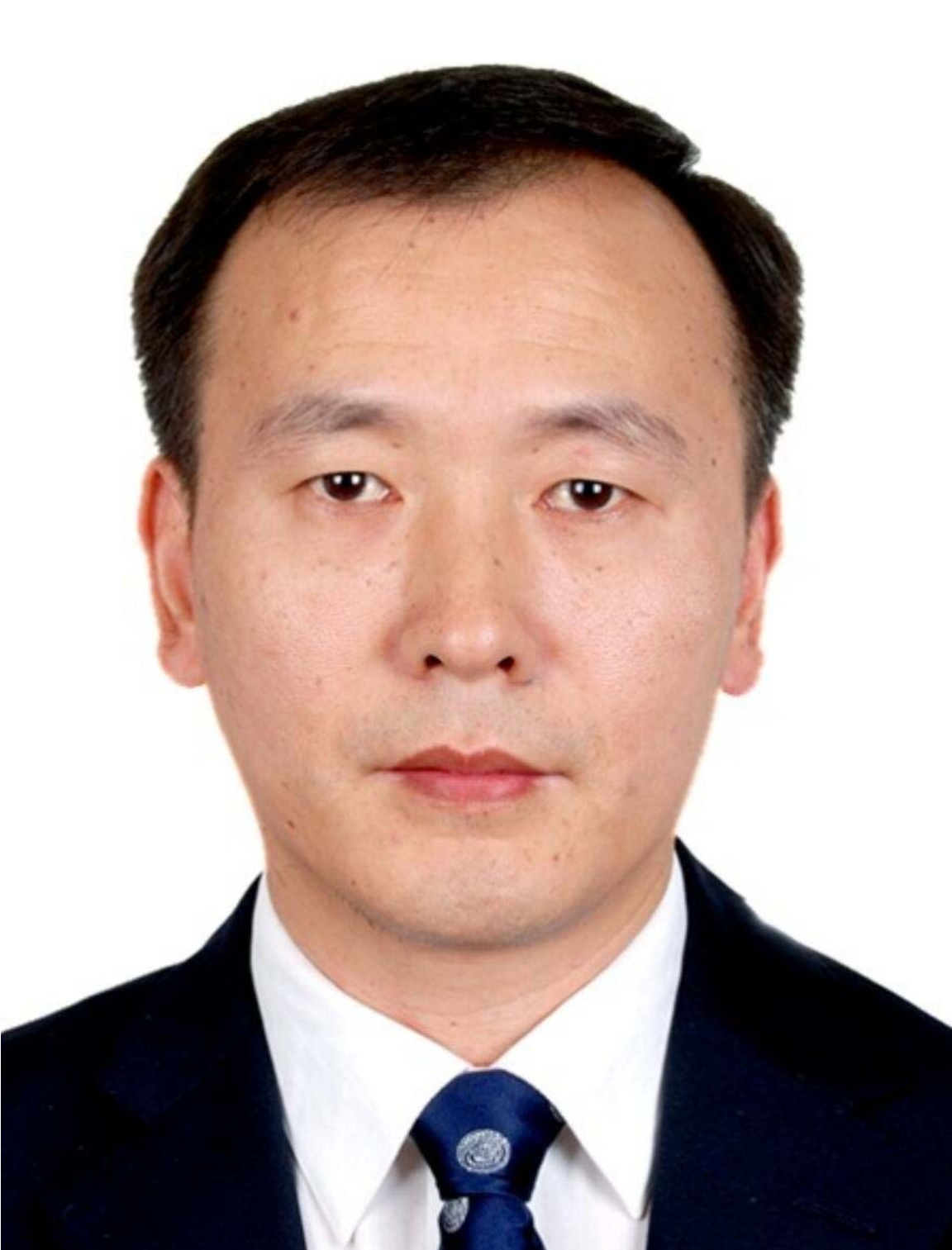}}]{Xiqi Gao}
	(S'92-AM'96-M'02-SM'07-F'15) received the Ph.D. degree in electrical engineering from Southeast University, Nanjing, China, in 1997. 
	
	Dr. Gao joined the Department of Radio Engineering, Southeast University, in April 1992. Since May 2001, he has been a professor of information systems and communications. From September 1999 to August 2000, he was a visiting scholar at Massachusetts Institute of Technology, Cambridge, MA, USA, and Boston University, Boston, MA. From August 2007 to July 2008, he visited the Darmstadt University of Technology, Darmstadt, Germany, as a Humboldt scholar. His current research interests include broadband multicarrier communications, MIMO wireless communications, channel estimation and turbo equalization, and multirate signal processing for wireless communications. From 2007 to 2012, he served as an Editor for the IEEE Transactions on Wireless Communications. From 2009 to 2013, he served as an Editor for the IEEE Transactions on Signal Processing. From 2015 to 2017, he served as an Editor for the IEEE Transactions on Communications. 
	
	Dr. Gao received the Science and Technology Awards of the State Education Ministry of China in 1998, 2006, and 2009, the National Technological Invention Award of China in 2011, and the 2011 IEEE Communications Society Stephen O. Rice Prize Paper Award in the field of communication theory.
\end{IEEEbiography}

\end{document}